%% file: main.tex
\documentclass[aps,prl,reprint,preprintnumbers,amsmath,amssymb,floatfix]{revtex4-1}

\usepackage{morefloats}
\usepackage[dvips]{graphicx}
\usepackage{color}
\usepackage{epsfig,graphicx,amsfonts,amsbsy}
\usepackage{amsmath,amsfonts,amsthm,amssymb}
\usepackage{appendix}
\usepackage{bbm}
\usepackage{makeidx}
\usepackage{url}
\usepackage{verbatim}
\usepackage[bookmarksnumbered,pdfpagelabels=true,plainpages=false,colorlinks=true,linkcolor=blue,citecolor=blue,urlcolor=blue]{hyperref}
\usepackage{array}
\usepackage{booktabs}
\usepackage{multirow}
\usepackage{bbm}
\usepackage{tabularx}
\usepackage{cancel,soul}
\usepackage{nicefrac, xfrac}

\makeatletter

\usepackage[version=3]{mhchem} 
\usepackage{feynmf}
\usepackage[retainorgcmds]{IEEEtrantools}
\usepackage{bbold}

\newcommand{\bs}{\boldsymbol{{\rm S}}}
\newcommand{\bh}{\boldsymbol{{\rm h}}}
\newcommand{\bt}{\boldsymbol{{\rm t}}}
\newcommand{\bH}{\boldsymbol{\mathcal{H}}} 
\newcommand{\bdot}{\boldsymbol{\cdot}} 
\newcommand{\btau}{\boldsymbol{{\rm \tau}}}
\newcommand{\br}{\boldsymbol{{\rm \mathcal{R}}}}
\newcommand{\bn}{\boldsymbol{{\rm N}}}
\newcommand{\ba}{\boldsymbol{{\rm a}}}
\newcommand{\bR}{\boldsymbol{{\rm r}}}
\newcommand{\bP}{\boldsymbol{{\rm p}}}
\newcommand{\bJ}{\boldsymbol{{\rm J}}}
\newcommand{\balpha}{\boldsymbol{{\rm \alpha}}}

\usepackage{xcolor}
\usepackage{nicefrac, xfrac}
\def \cedenna {Centro  de Nanociencia y Nanotecnología CEDENNA, Avda. Ecuador 3493, Santiago, Chile}
\def \fcfm {Departamento de F\'isica, FCFM, Universidad de Chile, Santiago, Chile.}
\def \puc {Instituto de F\'isica, Pontificia Universidad Cat\'olica de Chile, Casilla 306, Santiago, Chile}
\def \uaysen {Universidad de Aysen, Obispo Vielmo 62, Coyhaique, Chile}

\begin{document}


\title{Quantum spin fluctuations and the stability of atomically-sized Bloch points}
\author{Alonso Tapia$^1$}
\email{axtapia@uc.cl}
\author{Carlos Saji${}^{2,3}$}
\author{Alejandro Roldan${}^{4}$}
\author{Alvaro S. Nunez${}^{2,3}$}

\affiliation{${}^{1}$\puc}
\affiliation{${}^{2}$\fcfm}
\affiliation{${}^{3}$\cedenna}
\affiliation{${}^{4}$\uaysen}

\date{\today}
\begin{abstract}
We reveal the role of the spin variables' zero-point fluctuations (ZPFs) on the stability of Bloch point (BP) singularities. As topological solitons, BPs are important in topological transitions in nanomagnets. BPs present a singularity at their core, where the long-length-scale approximation fails. We found that ZPFs bloom nearby this core,  reducing the effective magnetic moment and increasing the BP's stability. As suggested by classical models, the magnonic eigenmodes found by our methods fit with the bound state of an electron surrounding a dyon, with a magnetic and an electric charge. 

\end{abstract}
\maketitle
\input{texts/Introduction}
\input{texts/BasicModel}
\input{texts/MagnonicRepresentation}
\input{texts/Results}
\input{texts/discussion}

\noindent {\it Acknowledgments.-} Funding is acknowledged from Fondecyt Regular 1230515, Fondecyt Iniciaci\'on 11201249 and Financiamiento Basal  para  Centros  Cient\'ificos  y  Tecnol\'ogicos  de  Excelencia AFB220001. Powered@NLHPC: This research was partially supported by the supercomputing infrastructure of the NLHPC, Chile (ECM-02).

%

\newpage
\input{texts/MicromagneticSimulations}

\input{texts/AppendixA}
\input{texts/AppendixB}
\input{texts/AppendixC}

\end{document}

%% file: texts/Introduction.tex
\noindent \emph{Introduction.-} Zero-point fluctuations of the electromagnetic field and their consequences\cite{Weinberg2015} are among the more startling and puzzling results in quantum-electrodynamics (QED)\cite{Coleman2018}. 
Zero-point fluctuations, or quantum fluctuations in the ground state of the electromagnetic field, are behind the remarkable Casimir effect\cite{Mostepanenko1997, Mohideen2014} and lie at the origin of the van der Waals interaction\cite{Abrikosov1975}. 
zero-point fluctuations were shown to affect the dynamics of real objects in the nanometric realm\cite{Lamoreaux1997}. Nowadays, they are on their way to becoming a starring member in the toolkit of nanoscience and nanotechnology\cite{Xu2022, Rodriguez2011}.

In the attempt to generalize zero-point fluctuations analysis to other science areas, a natural candidate is the magnetization field in magnetic systems\cite{Yuan20221}. The magnetization field, cast out of local spin moments, obeys a commutation algebra of the angular momentum. The lack of commutation between the different components of the magnetization leads in a direct way to pronounced zero-point fluctuations. Such fluctuations are particularly strong in the case of antiferromagnets\cite{Auerbach1994, Sachdev2011}; in many cases, they can even destroy the long-range magnetic order leading to exotic states known collectively as spin liquids\cite{Broholm2020}. 

Since the magnetization field commutes with the dominant contributions to the dynamics, zero-point fluctuations are drastically reduced. They become vanishingly small in the ideal situation of a Hamiltonian restricted to an isotropic Heisenberg exchange with a homogeneous ground state. There are two sources of quantum fluctuations. On the one hand, we have any term in the Hamiltonian not commuting with the magnetization order parameter. In this group, we list external fields, anisotropies of any kind, Dzyaloshinskii-Moriya interactions (DMI), and dipolar fields, among others. Complementarily, textures of the magnetization field will also induce quantum fluctuation leading to zero-point fluctuations effects. These include domain walls, vortices, helices, skyrmions, etc. 

The cases of  skyrmions\cite{Roldan2015} and helicoidal\cite{Roldan2014} states generated by DMIs have been studied. Due to the Hamiltonian and the texture induced, such calculations showed magnons displaying strong zero-point fluctuations. 
Those results become more relevant as the skyrmion size is reduced. They granted an entrance door into quantum skyrmionics\cite{Ochoa2019, Psaroudaki2022, Psaroudaki2017, Haller2022, Chen2021, Diaz2020, RoldanMolina2016}. 

In analogy with the electromagnetic case, where zero-point fluctuations lead to Casimir forces, it was claimed that zero-point fluctuations of the magnetization field lead to Casimir spin-torques (CSTs) acting over the magnetization texture. Analyzing magnetic interaction strengths might help elucidate the precise role of zero-point fluctuations in magnetization dynamics\cite{Bouaziz2020}. In \cite{Nakata2023}, it is argued that these CSTs can generate measurable effects in ferrimagnets, such as YIG, blazing the trail toward Casimir engineering. A key factor in identifying such magnon zero-point fluctuations is their dependence on an external magnetic field. Such dependence can distinguish them from other contributions, such as the photonic or phononic. 

In this letter, we study the effect of magnetization's zero-point fluctuations over a Bloch point. Bloch points, or singularities, are topological solitons of a three-dimensional magnetization field.\cite{Beg2019, Tejo2022, ZambranoRabanal2023, Saez2022, Li2020, Elas2011} Nearby the Bloch point's location, the spin configuration changes abruptly. In this way, one of the essential assumptions of micromagnetism fails to hold, and the magnetization field becomes ill-defined\cite{Dring1968, Galkina1993}. The presence of such singularities can change the behavior of a magnet deeply and controversially. 

\begin{figure}[ht]
\includegraphics[width=0.20\textwidth]{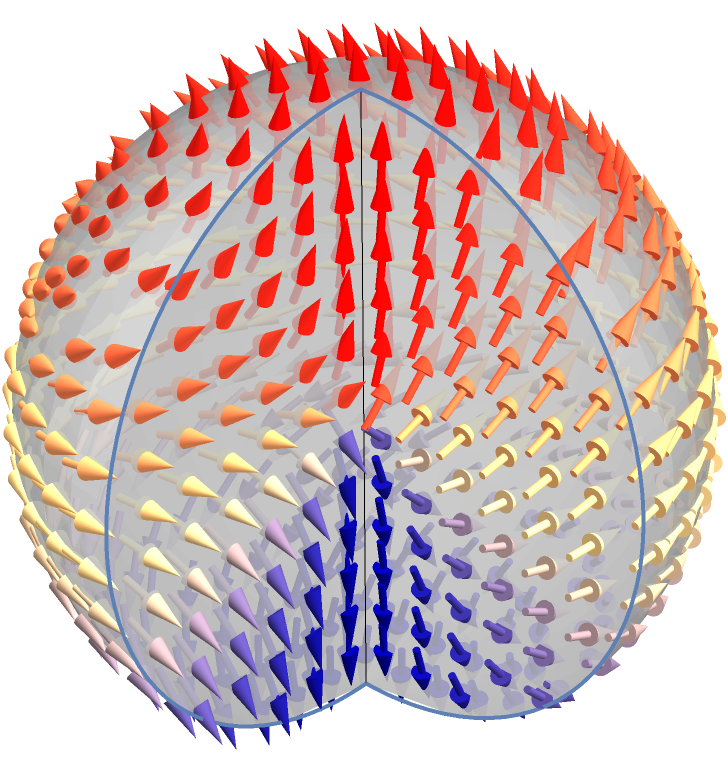}
\includegraphics[width=0.22\textwidth]{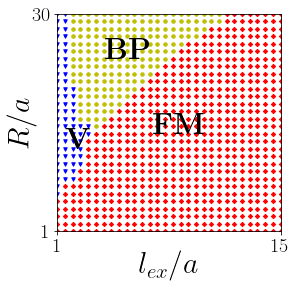}
\caption{(left panel) Cartoon of Bloch point stabilized through demagnetization field. In color, we have
highlighted the $z$-component of the magnetization
whose orientation is depicted by the arrows (right panel). Phase diagram of the system calculated through MuMax3 indicating the regions where the system becomes stable at a ferromagnetic (FM), vortex (V) or Bloch Point (BP) configuration as a function of material properties ($l_{ex}/a$) and geometry $R/a$. 
}
\label{fig: Bloch-Point cartoon}
\end{figure}

There are several contexts where such intricate configurations have been observed or, at least, strongly implied\cite{Izmozherov2019}. Slowly they are claiming an important role in modern magnetism. Experimental techniques such as X-Ray tomography play a major role in such a process, which probes the magnetization texture in the interior of the experimental samples\cite{Hierro-Rodriguez2020}. 
First proposed to play a role in the transition region  Bloch walls, numerical simulations provide a strong basis to expect them to appear in quasi-two-dimensional systems during the reversal of vortex cores. Recently, Bloch points played a role in topological transitions as sources and sinks of topological charge for unwinding a skyrmion crystal. A recent experimental work reports a static BP in cylindrical magnetic nanowires. 
Multiscale numerical algorithms\cite{Andreas2014} that treat the vicinity of the singularity and the long-distance regime with different paradigms seem to shed light on the subtle physics of BPs. However, a quantum perspective seems unavoidable near the singularity's core.
It was proved that the classical behavior of the linear perturbations (spin waves) around a singular Bloch point soliton is similar to that of an electron in a magnetic monopole field within a quantum system. Following such an analogy, the solution to this problem was analytically derived, enabling the determination of the spectrum and scattering of classical spin waves around a Bloch point field\cite{Elias2014, CarvalhoSantos2015}.
Our main result is that quantum spin waves, magnons, follow a similar pattern, and, what's more intriguing, they display zero-point fluctuations that are strongly at play in providing the Bloch point singularity stability. The quantized magnonic field fluctuates strongly near the origin, reducing the effective spin length. The highly frustrated neighborhood of the singularity sees, in this way, a dramatic reduction of its energy. 

%% file: texts/BasicModel.tex
\noindent \emph{Basic Model.-} Our study reduces the complexities of Bloch point energetics  to their bare minimum. We have a system of localized quantized spin moments (of length $S$) placed at the points of a BCC lattice with side $a$. We trim the lattice at the exterior, excluding all points beyond a fixed distance of $R$ from the origin. The origin is located at the center of the unit cell. In this way, no spin is located at the origin. The only interactions we include are the isotropic Heisenberg exchange, with strength $J$, among nearest neighbors and the dipolar field that connects each site to all the others with strength $Da^3$ times a dipolar factor with the usual form. In terms of the spin operators, denoting by $\bs_{i}$ the one corresponding to the site $i$, the Hamiltonian becomes:
\begin{equation}
\bH = -J\sum_{\langle i,j\rangle}\bs_{i} \bdot \bs_{j} + D a^3 \sum_{i<j}\left(\frac{\bs_{i}\bdot\bs_{j}- 3(\bs_{i}\bdot\hat{r}_{ij})(\bs_{j}\bdot\hat{r}_{ij})}{r_{ij}^{3}} \right),  \label{eq: Model Spin}
\end{equation}
where, as usual, $\langle i,j\rangle$ restricts the sum to nearest neighbors. The relative position of site $i$ with respect to site $j$
is labeled as $\vec{r}_{ij}=\vec{r}_i-\vec{r}_j$, that points along the direction $\hat{r}_{ij}$ and whose magnitude is $r_{ij}$. The four parameters of the model $R$, $a$, $J$, and $D$, are combined  into two dimensionless variables: $R/a$ and $\ell/a$, where $\ell$ is the exchange length ($\ell/a=\sqrt{\sfrac{J}{D}}$). Despite its simplicity, the model in Eq. (\ref{eq: Model Spin}) leads to a huge dimensional Hilbert space that turns intractable even within the spin wave approximation we will take later on. To circumvent this problem, we will address a reduced system using a scaling technique, tested\cite{Albuquerque2002} in the classical analog of Eq. (\ref{eq: Model Spin}). This analysis reduces the system spatial dimensions by $x^\eta$ (relative to a fixed lattice constant, $\eta\sim 0.5$) while the exchange energy $J$ is reduced by a factor $x$ accordingly. 

The model in Eq.(\ref{eq: Model Spin}) has been shown to exhibit BP singularities when addressed from the continuum perspective\cite{ZambranoRabanal2023, Tejo2022}. As is well known, Bloch points provide a colorful array of solutions in systems with only exchange energy. 
The effect of the dipolar coupling is to select specific solutions that reduce the magnetostatic energy out of that family. 

As a preliminary step, we found a classical solution, conforming a Bloch point, that minimizes (\ref{eq: Model Spin}) in the form $\bs^{\rm cl}_i=S\,\bn_i$, where the $\bn_{i} = (\cos\phi_{i}\sin\theta_{i}, \sin\phi_{i}\sin\theta_{i}, \cos\theta_{i})$ are classical unit vectors.  We studied the stability of such a texture, purely on classical grounds, in terms of the different system parameters. The results of such analysis are presented in Fig. (\ref{fig: Bloch-Point cartoon}) right panel.

%% file: texts/MagnonicRepresentation.tex
\noindent \emph{Magnonic representation and expansion.-} Starting from the classical solution  $\bs^{\rm cl}_i=S\,\bn_i$, we return to the quantum Hamiltonian and perform the Holstein-Primakoff expansion\cite{Auerbach1994,Holstein1940}. In this representation, the quantum spins are mapped into a system of harmonic oscillators with ladder operators $\ba^{\dagger}_{i}$ and $\ba_{i}$.  We keep terms up to quadratic order in the ladder operators to avoid an intractable situation. The physical consequence of this approximation is to confine the physics to non-interacting magnons. This approximation will remain valid as long as $S\gg \langle\ba^\dagger_i\ba_i\rangle$, which will be tested afterward for consistency. 
This transformation reduces the complete Hamiltonian in Eq. (\ref{eq: Model Spin}) to a quadratic form in the bosonic operators $\ba^{\dagger}_{i}$ and $\ba_{i}$.
\begin{equation}
    \bH = \mathcal{E}_{cl} + \mathcal{E}_{0} + \sum_{i,j}(\bt_{ij}\,\ba^{\dagger}_{i}\ba_{j}+\btau_{ij}\,\ba^{\dagger}_{i}\ba^{\dagger}_{j}+h.c.)
    \label{eq: Hamiltonian boson}
\end{equation}
the first term corresponds to the classical energy that arises from replacing $\bs^{\rm cl}_i$ into Eq. (\ref{eq: Model Spin}), and the second term is a residual contribution that is accumulated by bringing the bosonic expansion into normal ordering. Finally, we have the quadratic Hamiltonian in terms of the operators $\ba^{\dagger}_{i}$ and $\ba_{i}$. 
We note the presence of anomalous terms, proportional to $\btau_{ij}$, that break the Hamiltonian's $U(1)$ symmetry. These arise from the dipolar interaction and the lack of collinearity of the classical configuration. Such terms explicitly break the magnon number conservation and are a source of zero-point fluctuations. Due to the dipolar interaction, the couplings in Hamiltonian in Eq.(\ref{eq: Hamiltonian boson}) are long-ranged. However, they can be readily calculated starting from a classical solution, as Appendix A shows.

We bring Eq. (\ref{eq: Hamiltonian boson}) into diagonal form\cite{Bogoljubov1958, Wagner1986, Kittel1987, Colpa1978}. The resulting Hamiltonian acquires the familiar form of a collection of quantum harmonic oscillators, corresponding to the magnon states, labeled by the index $\nu$, each with a different frequency, $\varepsilon_\nu$, the normal mode energies. Each state is characterized by a bosonic operator $\balpha^{\dagger}_{\nu}$. 
In terms of those operators, the Hamiltonian reads:
\begin{equation}
\bH = \mathcal{E}_{cl} + \mathcal{E}_{0} + \sum_{\nu=1}^{N}\varepsilon_{\nu}\left(\balpha^{\dagger}_{\nu}\balpha_{\nu} + \frac{1}{2}\right).
\end{equation}
The state regarded as the vacuum of excitations is annihilated by each one of the lowering operators $\balpha_{\nu}|\rm{vac}\rangle=0$. The expectation value of the energy on the vacuum state $\langle \rm{vac}|\bH|\rm{vac}\rangle$, differs from the classical value by $\mathcal{E}_{ZP}=\mathcal{E}_{0} +\sum_{\nu=1}^{N}\varepsilon_{\nu}/2$, that corresponds to  the magnon zero-point energy.

%% file: texts/Results.tex
\begin{figure}[ht]
    \centering
    \includegraphics[width=0.23\textwidth]{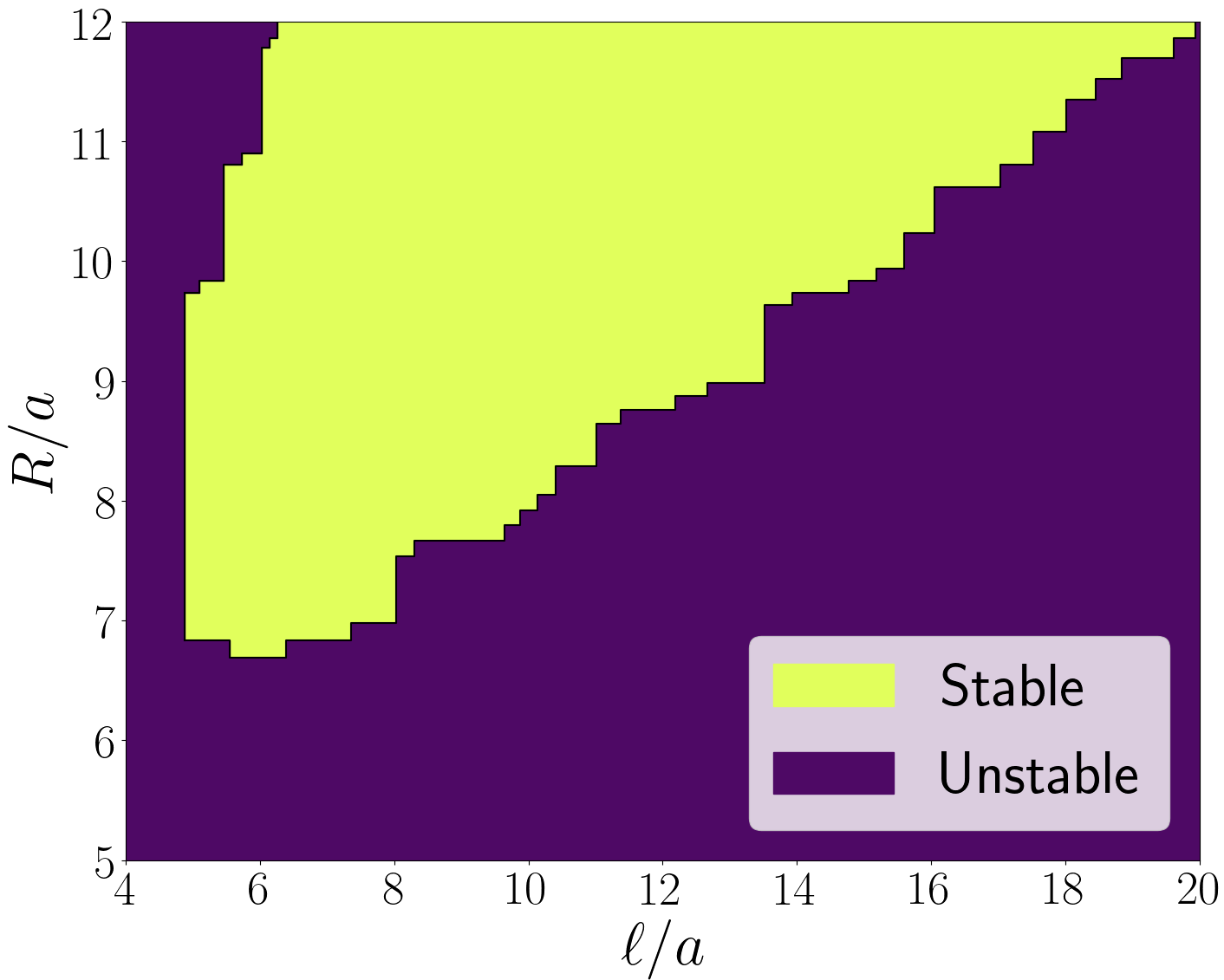}
    \includegraphics[width=0.23\textwidth]{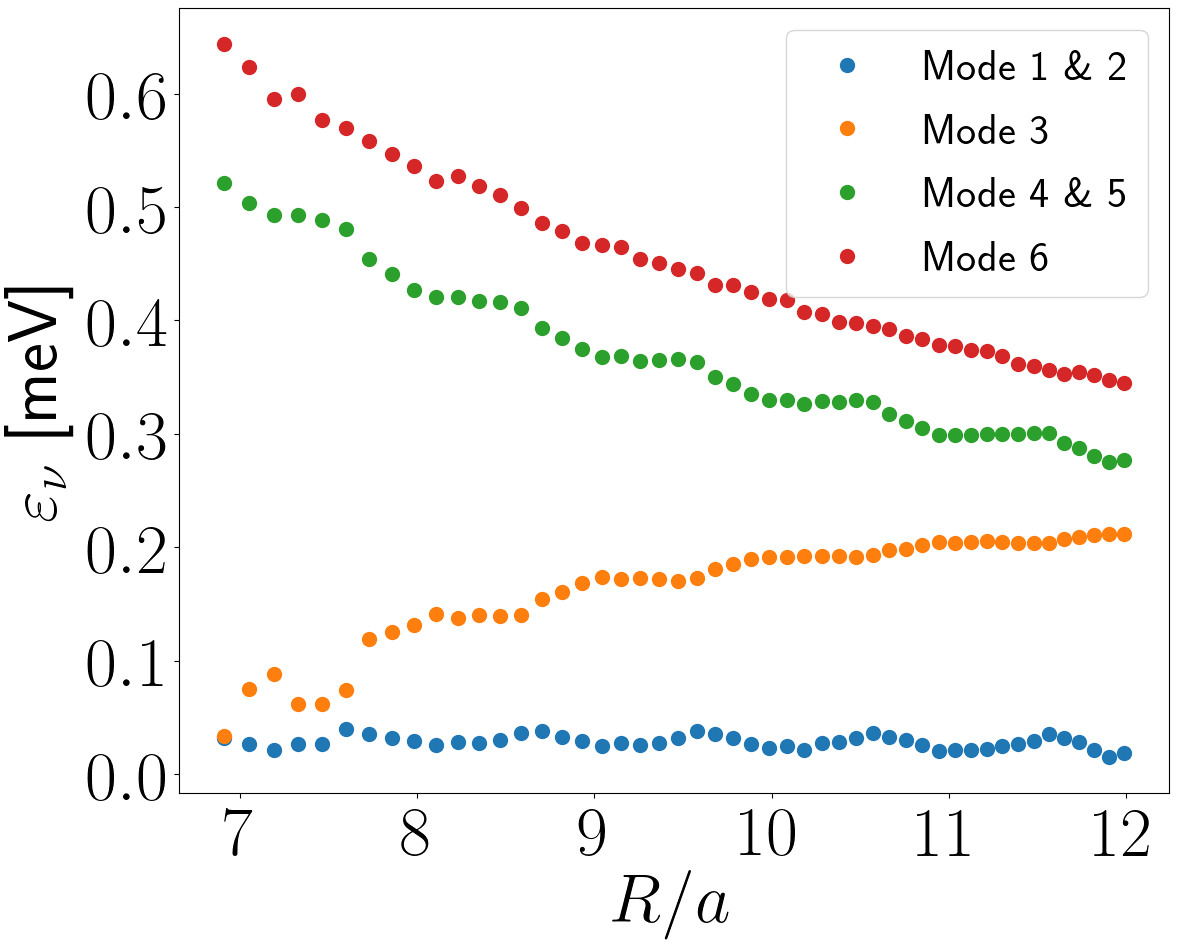}
    \includegraphics[width=0.23\textwidth]{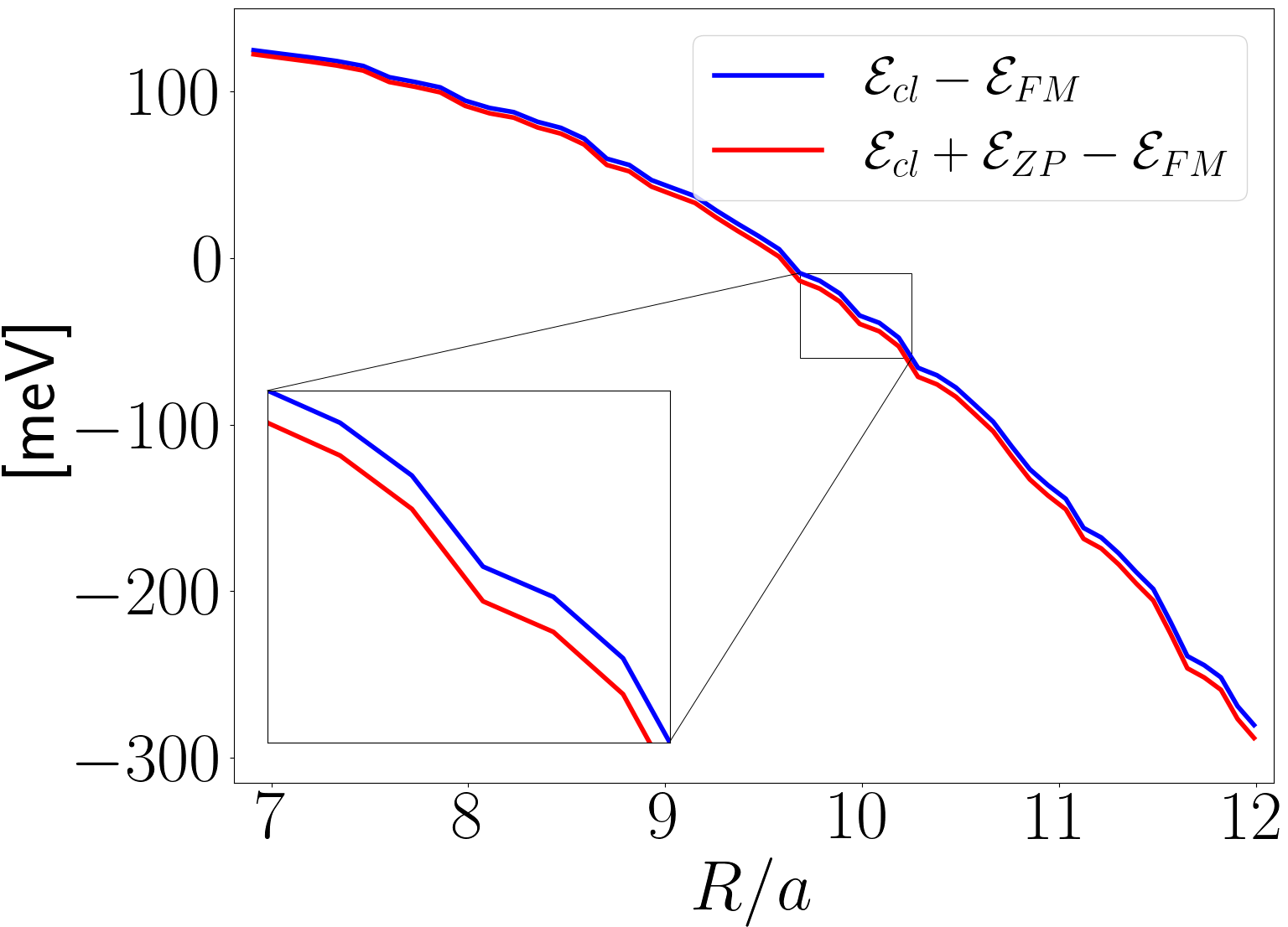}
    \includegraphics[width=0.23\textwidth]{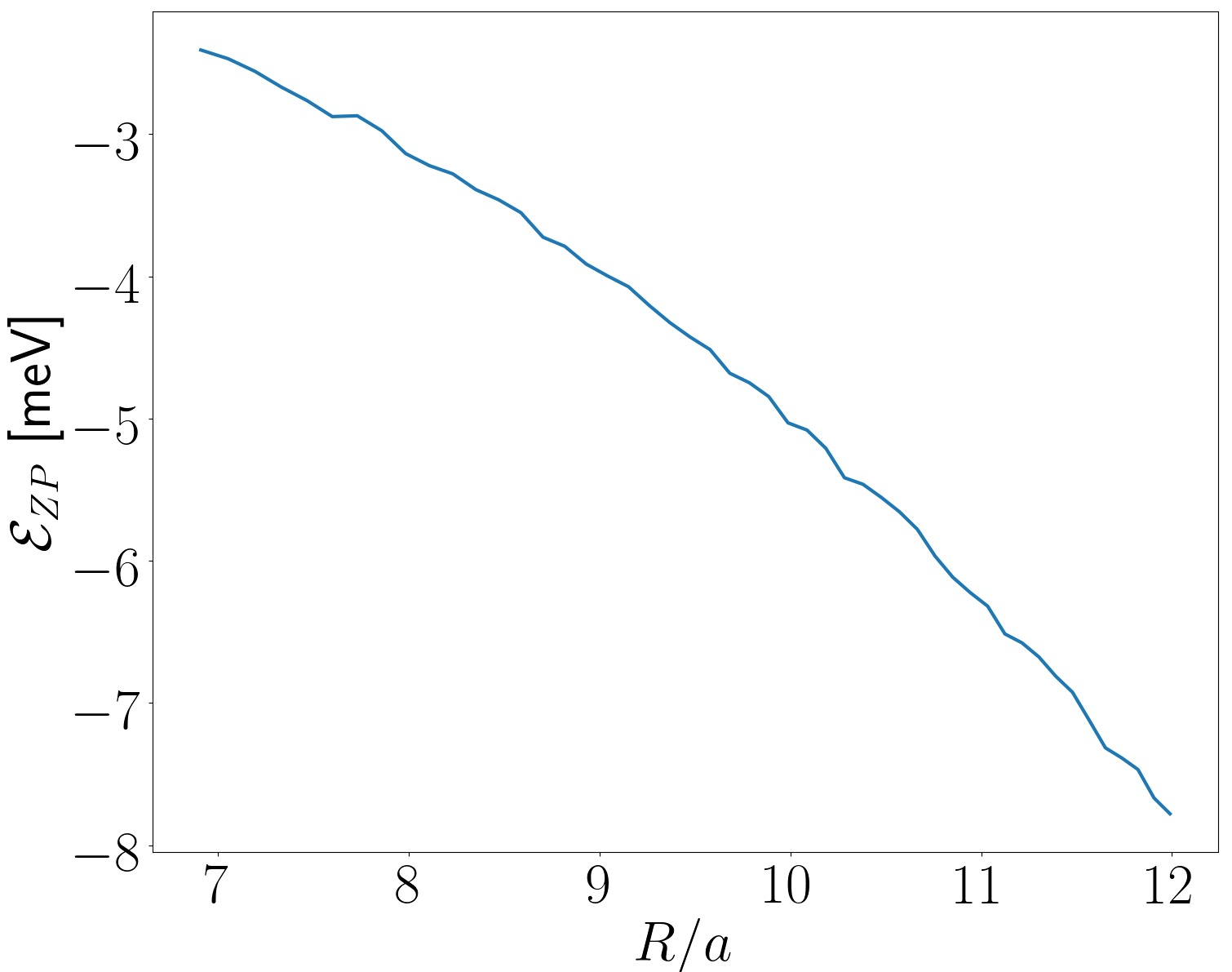}
    \caption{(top left panel) Phase diagram summarizing the stability analysis performed over the plane $R/a$ vs. $\ell/a$.  (top right panel) The energy eigenvalue of the first six magnonic states excited as a function of the radius is displayed in this panel.  (bottom left panel) The comparison between the classical energy of the system and the corrected energy by $\mathcal{E}_{ZP}$, as a function of the radius, plotted with respect to the energy of the ferromagnetic state ($\mathcal{E}_{FM}$). There is a lowering of the overall energy as a consequence of quantum fluctuations. (bottom right panel) This figure shows the zero-point energy of the system as a function of system size. }
    \label{fig: phase diagram}
\end{figure}

\noindent\emph{Results.-}
Our first result stems from a stability analysis. We have identified a region in parameter space ($\ell/a$ vs. $R/a$) where the classically stable Bloch points are also stable under the effect of quantum fluctuations. This region can be observed in Fig. (\ref{fig: phase diagram}). We identify three regions. The center region highlights the geometric and magnetic parameters stabilizing the Bloch point.  The stable area is identified by having only positive eigenvalues of the dynamic matrix in Eq. (\ref{eq: Hamiltonian boson}). This allows for correctly identifying the magnonic vacuum and constructing an unambiguous magnon Hilbert space. The unstable regions have one or more negative eigenvalues of the same matrix.  From now onward, we will focus only on the part where the Bloch points are stable.

Following our analysis, we calculate the magnon frequencies. The magnetic sphere acts as a magnon cavity within which the magnons are confined. Magnons spread across the sphere, and the effects of the confining edges of the system quantize their frequencies.  These boundary conditions reflect in their degeneracy structure information on the symmetry of the underlying magnetic texture. We present our results in the top right panel of Fig.(\ref{fig: phase diagram}). It is important to notice that there are degenerate states, which means that different states can have the same energy level. 
Once the frequencies are known, providing a value for the zero-point energy is simple. This constitutes the bottom panel of Fig.(\ref{fig: phase diagram}).

\begin{figure}[ht]
    \centering
    \includegraphics[width=0.21\textwidth]{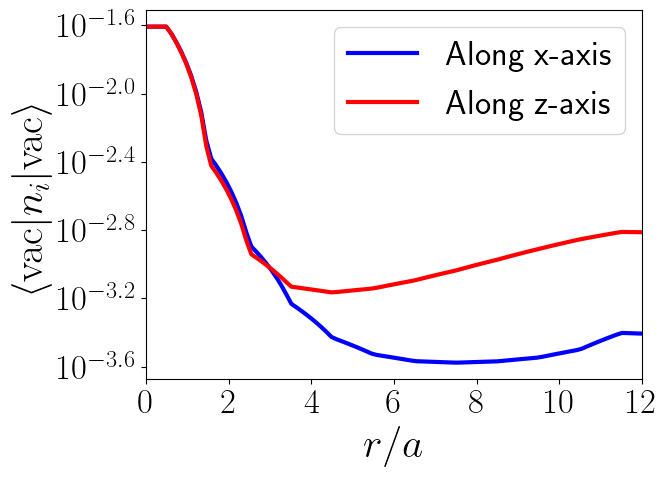}
    \includegraphics[width=0.21\textwidth]{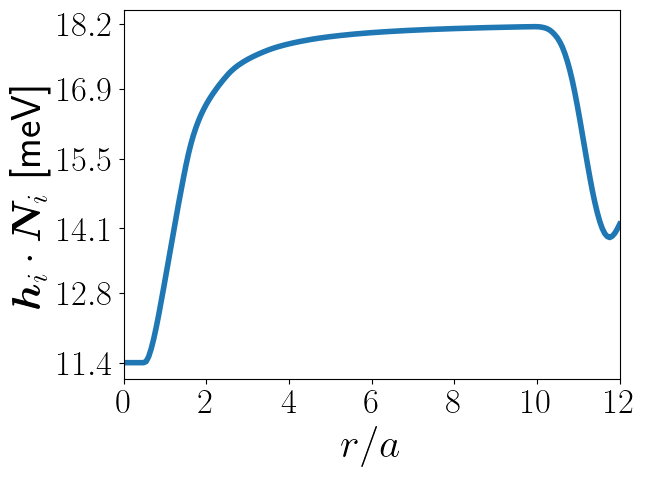}
    \includegraphics[width=0.21\textwidth]{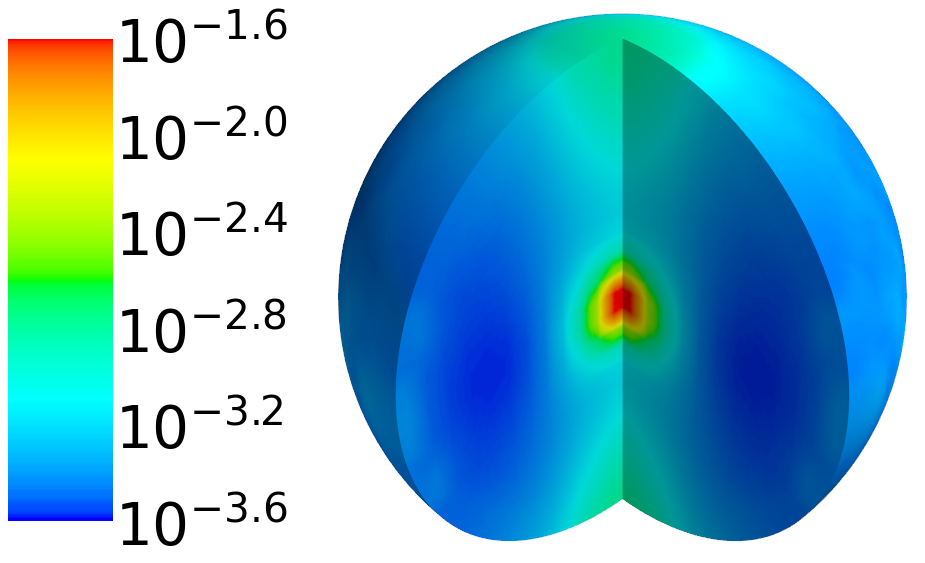}
    \includegraphics[width=0.21\textwidth]{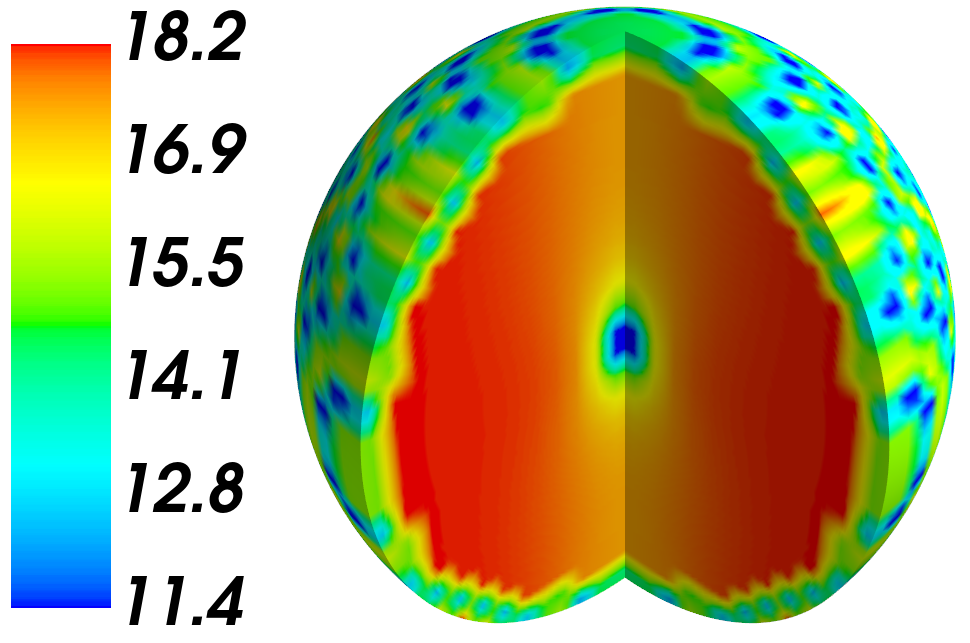}
    \caption{ (left panel, top and bottom) Strength of the zero-point fluctuations, $\left\langle {\rm vac}|n_{i}|{\rm vac}\right\rangle$. We display the logarithm of this quantity using a color scheme alongside the corresponding figure. (right panel, top and bottom) We calculate the CSFs. We show its component along the classical vector spin. }
    \label{fig: ZP fluctuations}
\end{figure}

The strength of the ground state fluctuations of the system is quantified through the expectation value of the occupation number operator $n_i=\langle {\rm vac}|\ba^\dagger_i\ba_i|{\rm vac}\rangle$. After performing the diagonalization, we evaluated these expectation values. They are displayed in the left panel of Fig.(\ref{fig: ZP fluctuations}). We see rather conclusively that most fluctuations occur in the region nearby the singularity. A tenuous cloud of fluctuations escapes the origin in the directions of the poles, as can be seen in Fig. (\ref{fig: ZP fluctuations}). Since mostly collinear spins characterize these regions, the origin of such fluctuations lies in the dipolar field.  

A most intriguing physical characterization of the ZPE is performed through the calculation of Casimir's spin field (CSF):
$\bh_i=-\frac{\delta \mathcal{E}_{ZP}}{\delta\bn_i}.$
We evaluate it and Casimir's spin torque (CST, $\btau_i=\bn_i\times\bh_i$). We show the projection of the CSF along the spin direction in Fig. (\ref{fig: ZP fluctuations}), where it can be appreciated that it acts as a stabilizing influence throughout the system.

\begin{figure}[ht]
    \centering
    \includegraphics[width=0.5\textwidth]{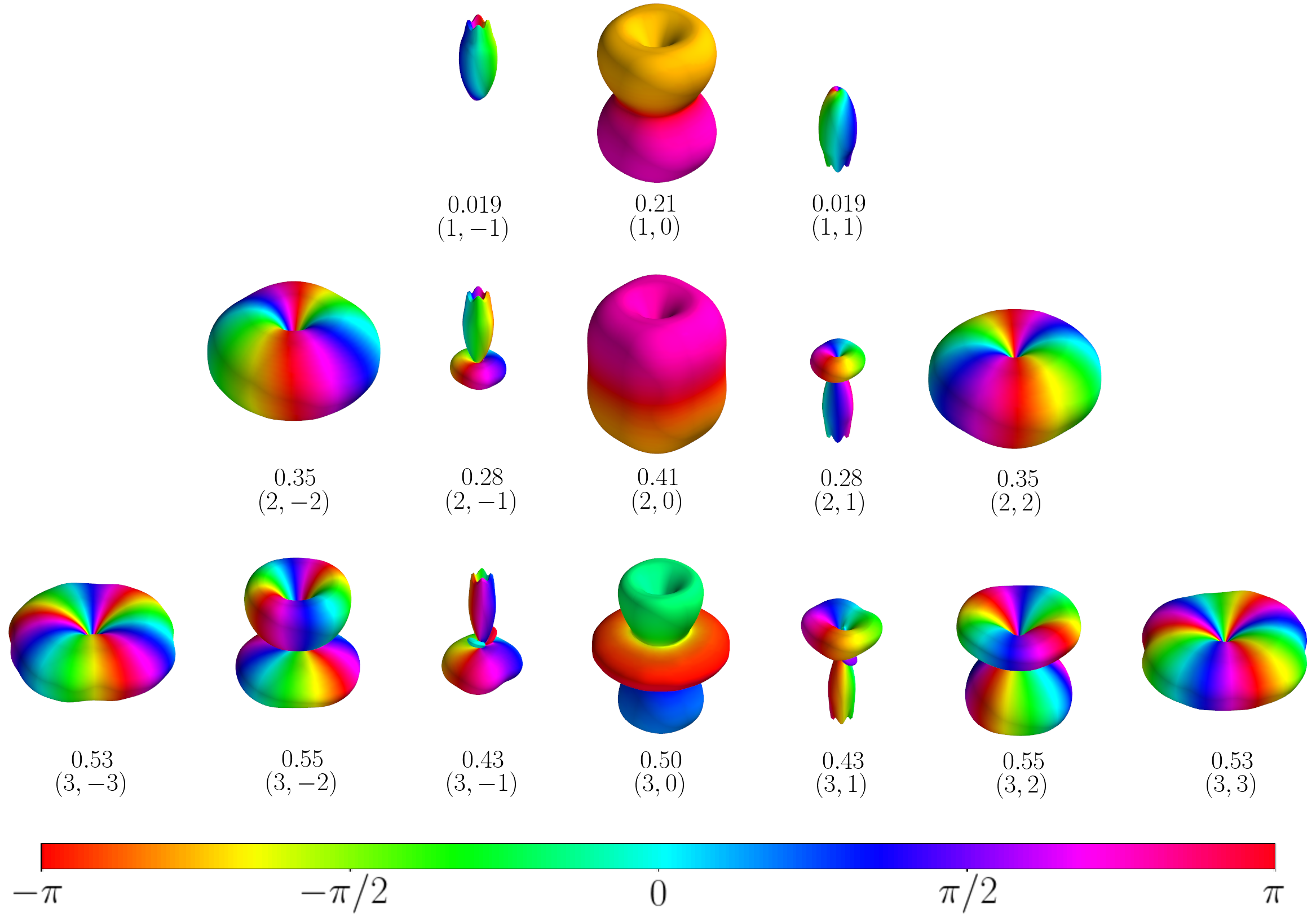}
    \caption{$\Psi^\nu$ at radius $9.5\,a$, for the first 15 modes, in color the phase and in distance the module. With the energy value in meV and the quantum numbers $(l,m)$. As can be appreciated, the different modes $\nu$ display different values of the angular momentum.}
    \label{fig: wave function}
\end{figure}

Finally, we have calculated a wave-function-like quantity that characterized the local strength and phase of the magnonic disturbances that constitute the eigenmodes of the system. We have defined the wave-function of the magnon mode $\nu$ at site $i$ as 
$\Psi^\nu_i=\langle {\rm vac}|\ba_i\balpha^\dagger_\nu|{\rm vac}\rangle$.
We have represented our results graphically in Fig.(\ref{fig: wave function}). It is apparent from those results that the magnons behave as eigenfunctions of the generalized angular momentum operator $\bJ=\bR\times\bP+\hat{\bR}$ as expected for a charged particle undergoing an orbit around a magnetic monopole of unit magnetic charge\cite{Schwinger1976}. The additional $\hat{\bR}$ in the generalized angular momentum arises from the electromagnetic angular momentum stemming from the monopole magnetic field and electric field associated with the coupling between the magnon and the monopole. The notion that magnons around a BP were analogous to a particle on a magnetic monopole was put forward in \cite{Elias2014}.  

%% file: texts/discussion.tex
\noindent \emph{Discussion.-} %
The confined magnons in a Bloch point configuration can exhibit a variety of interesting phenomena, such as zero-point fluctuations and the experience of emergent $U(1)$ magnetic monopoles\cite{Manton2004, Weinberg2012}. The strong zero-point fluctuations surrounding the BP singularity are the key to providing the stability these structures require. The effective reduction of the spin effective length is a stabilizing mechanism often quoted in the literature. By the same token, understanding this reduction is a key aspect of controlling topological transitions mediated by Bloch points singularities\cite{Birch2021}.  

The emergent magnetic monopoles have been reported as arising solely from exchange interactions in extended systems without dipolar coupling\cite{Elias2014, CarvalhoSantos2015}. The emergence of magnetic monopoles in confined magnons can be regarded as a new way to manipulate magnetic information and opens up new possibilities for magnonic devices in a non-local and topological fashion.
 These magnetic monopoles can act as sources or sinks of an effective magnetic flux felt by the magnons. The net emergent magnetic charge enclosed by the monopoles can be measured experimentally by non-local interferometry experiments\cite{Li2018} and can be a game-changing tool in magnon-magnon entanglement experiments\cite{Zhang2019, Yuan2020, Azimi2021, Ren2022}. Our report provides clear theoretical evidence that this is the case even at the innermost atomic scale vicinity of the Bloch point singularity. The ability to control and manipulate magnetic monopoles can lead to the development of new magnetic storage devices and logic elements, which have higher data densities and faster processing speeds than existing technologies\cite{Yuan20221}. 

In conclusion, by studying confined magnons on Bloch points singularities, we have provided exciting new insights into the physics of zero-point fluctuations and emergent magnetic monopoles. Zero-point fluctuations stabilize the structure, and the magnetic monopole dictates the basic structure underlying the magnonic spectra. The ability to control and manipulate magnetic zero-point fluctuations and emergent magnetic monopoles will likely lead to faster, more efficient, and higher-density magnonic devices.

%% file: texts/MicromagneticSimulations.tex
\section{Micromagnetics Simulations}
In this study, we employed the GPU-accelerated Mumax3\cite{Vansteenkiste2014} to perform numerical simulations on a system with a spherical geometry of radius $R$ ranging from $2\ [\mathrm{nm}]$ to $40\ [\mathrm{nm}]$. The goal was to investigate the behavior of the system under varying exchange stiffness $A_{ex}$, which was chosen to span from $1\ [\mathrm{pJ/m}]$ to $140\ [\mathrm{pJ/m}]$. We calculated the exchange length $l_{ex}$ as $l_{ex}=\sqrt{2 A_{ex}/(\mu_{0}M_{s}^{2})}$, with a saturation magnetization $M_{s}=10^{6}\mathrm{[A/m]}$, which helped us to analyze the system's behavior further. We carefully controlled these parameters to ensure they accurately modeled the system's behavior while remaining within a realistic range. Our findings provide insight into the magnetic properties of the system under these varying conditions and can aid in the development of future technologies that rely on magnetic materials.
\begin{figure}[ht]
    \centering
    \includegraphics[width=0.3\textwidth]{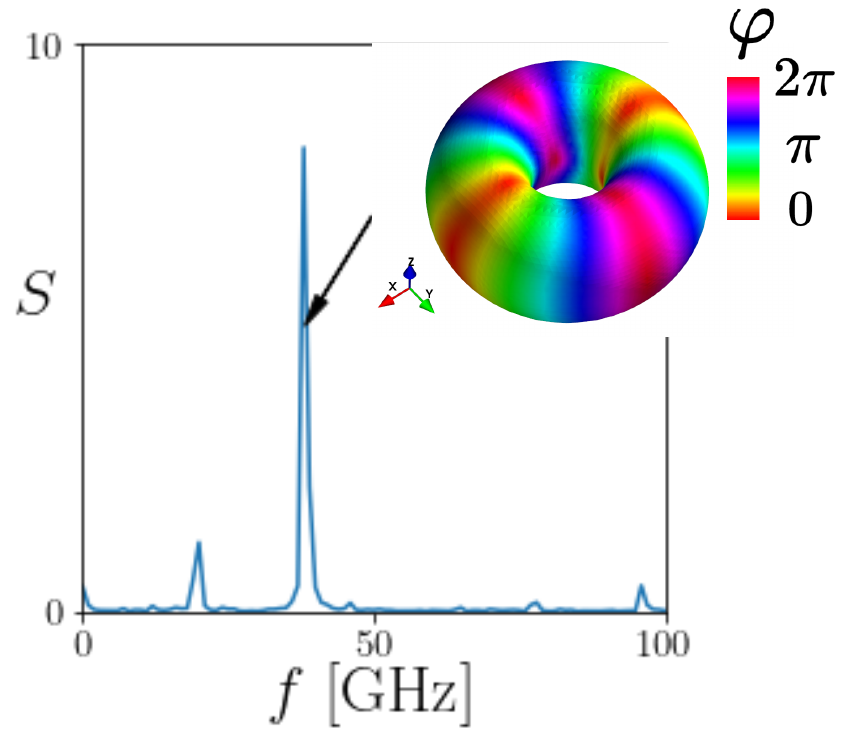}
    \caption{ Power Spectrum of a Bloch Point with $l_{\mathrm{ex}}=1.8\ [\mathrm{nm}]$ and of radius $R=30\ [\mathrm{nm}]$, calculated by means of the Fast Fourier Transform of the spin waves: $S(f)=\sum_{j} |\delta \boldsymbol{n}_{j}(f)|^{2}$ excited by a magnetic field given by $\boldsymbol{B}_{ac}(t)=\mathrm{sinc}(100[\mathrm{GHz}] t)[\mathrm{mT}]\ \hat{\boldsymbol{x}}$. The inset figure depicts the spatial amplitude of the corresponding resonance mode, where $\varphi$  is the complex phase of $\delta n_{j}^{x}(f)+i\delta n_{j}^{y}(f)$. }
    \label{fig: Power Spectrum}
\end{figure}

%% file: texts/AppendixA.tex
\section{Coefficients in the quadratic Hamiltonian}
The Holstein-Primakoff mapping is: 
\begin{eqnarray}
\bs_{i}\cdot\bn_{i}=S-n_{i} \\
\bs^{+}_{i}=\sqrt{2S-n_{i}}\,\ba_{i}\approx\sqrt{2S}\ba_{i} \nonumber\\
\bs_{i}^{-}=\ba^{\dagger}_{i}\sqrt{2S-n_{i}}\approx\sqrt{2S}\ba^{\dagger}_{i}\nonumber
\end{eqnarray}
where $n_i=\ba^\dagger_i\ba_i$.

The expansion occurs on a local basis whose $z$ axis is oriented along $\bn_i$. To represent the dynamics on a global basis, we need to rotate the operators in the following form:
\begin{equation}
\bs_i = 
\br(\bn_i)
\left(
\begin{matrix}
\sqrt{\frac{S}{2}}(\ba^{\dagger}_{i}+\ba_{i})\\
i\sqrt{\frac{S}{2}}(\ba^{\dagger}_{i}-\ba_{i})\\
S-\ba^{\dagger}_{i}\ba_{i}
\end{matrix}
\right)
\end{equation}
where $\br(\bn)$ is defined below.
As an auxiliary measure, we define before evaluating the matrix elements of $\bt$ and $\btau$.
$\vec{u}_{ij}  = \br^{t}(\bn_{i})\hat{r}_{ij}$
and 
$\vec{v}_{ij} = \br^{t}(\bn_{j})\hat{r}_{ij}$.
In the latter expressions, $\br(\bn)$ is the rotation matrix that brings $\bn$ toward the $\hat{z}$ axis. In terms of the spherical angles:
$$
\br(\bn) =
\left(
\begin{matrix}
\cos\theta\cos\phi & -\sin\phi & \sin\theta\cos\phi\\
\cos\theta\sin\phi & \cos\phi & \sin\theta\sin\phi\\
-\sin\theta & 0 & \cos\theta
\end{matrix}
\right).
$$
Eq. (\ref{eq: Hamiltonian boson}) can be written as 
$$\bH= \mathcal{E}_{cl} + \mathcal{E}_{0} + \ba^{\dagger}\mathcal{D}\ba,$$
where
$$
\ba^{\dagger} = \left(\ba_{1}^{\dagger}, \cdots, \ba_{N}^{\dagger},\ba_{1},\cdots,\ba_{N}\right),
$$
and $N$ is the number of sites in the system.
The dynamical matrix $\mathcal{D}$,
\begin{equation}
\mathcal{D} = \left(
\begin{matrix}
\bt & \btau\\
\btau^{*} & \bt^{*}
\end{matrix}
\right).
\end{equation}
is a $(2N)\times(2N)$ matrix whose elements depend on the magnetic configuration of the classical state and on the geometry of the system.

After a lengthy but straightforward calculation can evaluate the matrix elements of $\mathcal{D}$. 
The diagonal matrix elements are:
\begin{widetext}
\begin{eqnarray}
    \bt_{ii} = \frac{SJ}{2}\sum_{j  \langle i,j\rangle}\bn_{i}\cdot\bn_{j} - \frac{SDa^3}{2}\sum_{j\neq i}\left(\frac{\bn_{i}\cdot\bn_{j} - 3(\bn_{i}\cdot\hat{r}_{ij})(\bn_{j}\cdot\hat{r}_{ij})}{r_{ij}^{3}}\right)
\end{eqnarray}

$$\btau_{ii} = 0$$

On the other hand, in the case of nearest neighbors:

\begin{eqnarray}
\bt_{ij} &=& \frac{S}{4}\left(\frac{Da^3}{r_{ij}^{3}}-J\right)\left[(1+\cos\theta_{i}\cos\theta_{j})\cos(\phi_{i}-\phi_{j})+\sin\theta_{i}\sin\theta_{j}-i\sin(\phi_{i}-\phi_{j})(\cos\theta_{i}+\cos\theta_{j})\right] \nonumber\\&-& \frac{3SDa^3}{4r_{ij}^{3}}\left[u_{ij}^{x}v_{ij}^{x}+u_{ij}^{y}v_{ij}^{y}+i\left(u_{ij}^{y}v_{ij}^{x}-u_{ij}^{x}v_{ij}^{y}\right)\right]
\end{eqnarray}

\begin{eqnarray}
\btau_{ij} &=& \frac{S}{4}\left(J-\frac{Da^3}{r_{ij}^{3}}\right)\left[(1-\cos\theta_{i}\cos\theta_{j})\cos(\phi_{i}-\phi_{j})-\sin\theta_{i}\sin\theta_{j}+i\sin(\phi_{i}-\phi_{j})(\cos\theta_{i}-\cos\theta_{j})\right] \nonumber\\&-& \frac{3SDa^3}{4r_{ij}^{3}}\left[u_{ij}^{x}v_{ij}^{x}-u_{ij}^{y}v_{ij}^{y}+i\left(u_{ij}^{y}v_{ij}^{x}+u_{ij}^{x}v_{ij}^{y}\right)\right]
\end{eqnarray}

The rest of the matrix elements, for an arbitrary pair, can be written as:

\begin{eqnarray}
\bt_{ij} &=& \frac{SDa^3}{4r_{ij}^{3}}\left[(1+\cos\theta_{i}\cos\theta_{j})\cos(\phi_{i}-\phi_{j})+\sin\theta_{i}\sin\theta_{j}-i\sin(\phi_{i}-\phi_{j})(\cos\theta_{i}+\cos\theta_{j})\right]\nonumber\\ &-& \frac{3SDa^3}{4r_{ij}^{3}}\left[u_{ij}^{x}v_{ij}^{x}+u_{ij}^{y}v_{ij}^{y}+i\left(u_{ij}^{y}v_{ij}^{x}-u_{ij}^{x}v_{ij}^{y}\right)\right]
\end{eqnarray}

\begin{eqnarray}
\btau_{ij} &=& -\frac{SDa^3}{4r_{ij}^{3}}\left[(1-\cos\theta_{i}\cos\theta_{j})\cos(\phi_{i}-\phi_{j})-\sin\theta_{i}\sin\theta_{j}+i\sin(\phi_{i}-\phi_{j})(\cos\theta_{i}-\cos\theta_{j})\right]\nonumber\\&-& \frac{3SDa^3}{4r_{ij}^{3}}\left[u_{ij}^{x}v_{ij}^{x}-u_{ij}^{y}v_{ij}^{y}+i\left(u_{ij}^{y}v_{ij}^{x}+u_{ij}^{x}v_{ij}^{y}\right)\right]
\end{eqnarray}

\end{widetext}

%% file: texts/AppendixB.tex
\newpage

\section{Eigenmodes}

\begin{figure}[h!]
    \centering
    \includegraphics[width=0.46\textwidth]{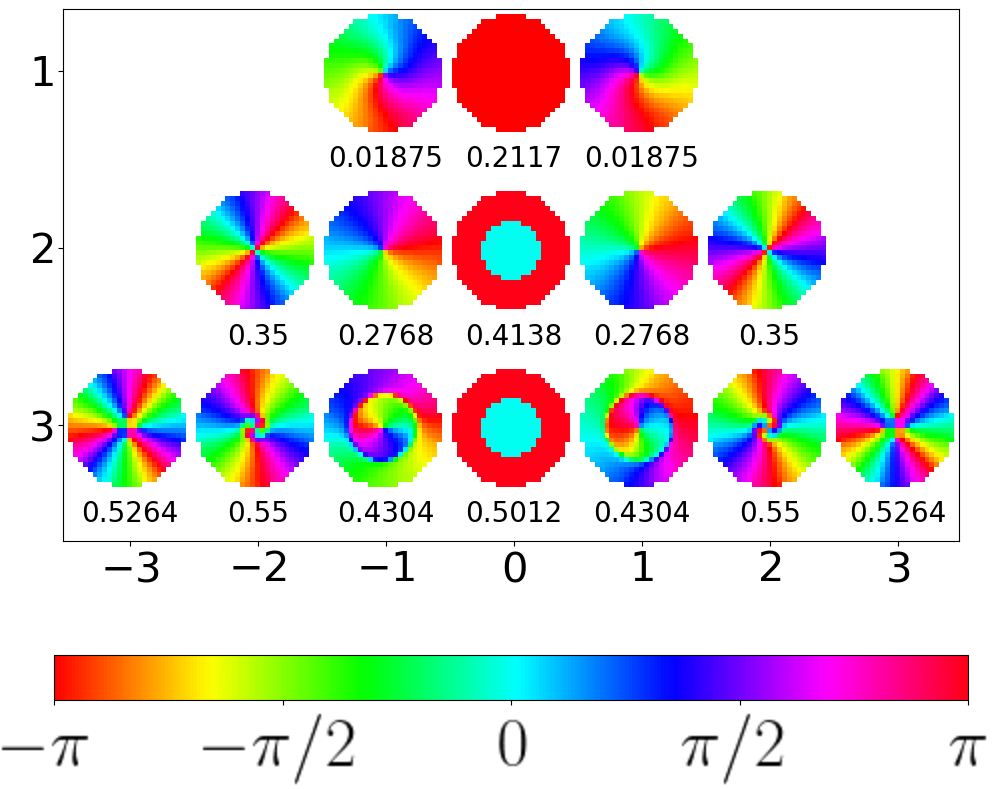}
    \caption{ Phase of $\Psi^\nu$ along the equatorial plane of the system, ordered according to the quantum numbers $l$ and $m$, with the energy value in meV.}
    \label{fig: wave function XY}
\end{figure}

\newpage

\begin{figure}[ht]
    \centering
    \includegraphics[width=0.48\textwidth]{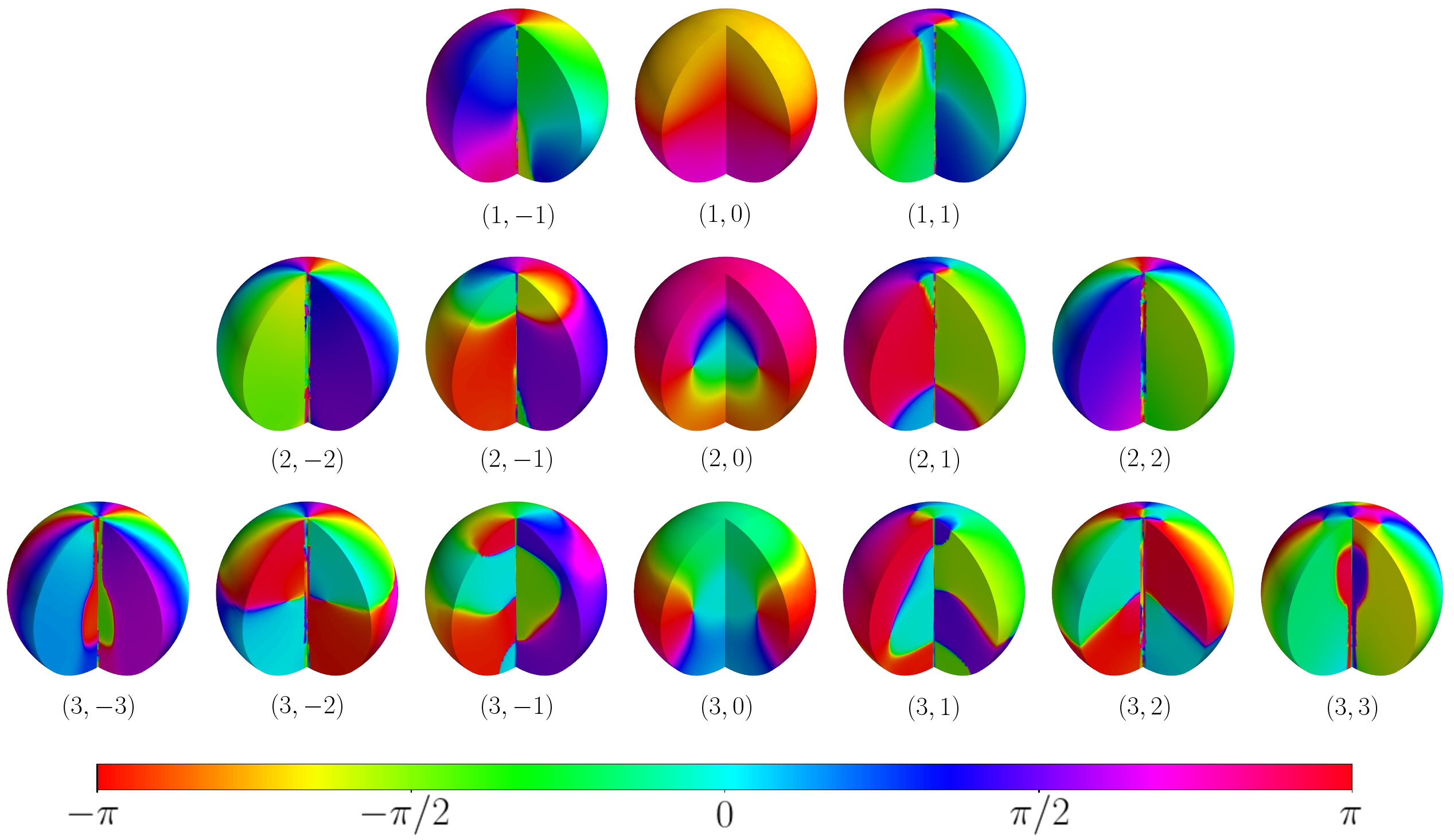}
    \caption{ Phase of $\Psi^\nu$, inside the sphere, with the quantum numbers $(l,m)$.}
    \label{fig: wave function sphere}
\end{figure}

\newpage

%% file: texts/AppendixC.tex
\newpage

\section{Casimir's Spin Torque}

\begin{figure}[h!]
\includegraphics[width=0.22\textwidth]{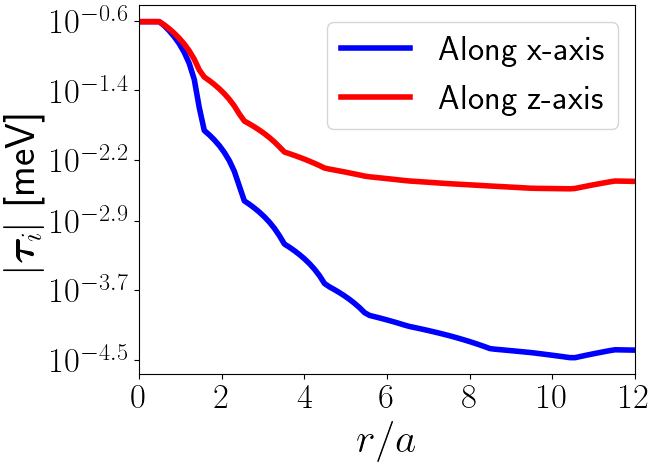}
\includegraphics[width=0.25\textwidth]{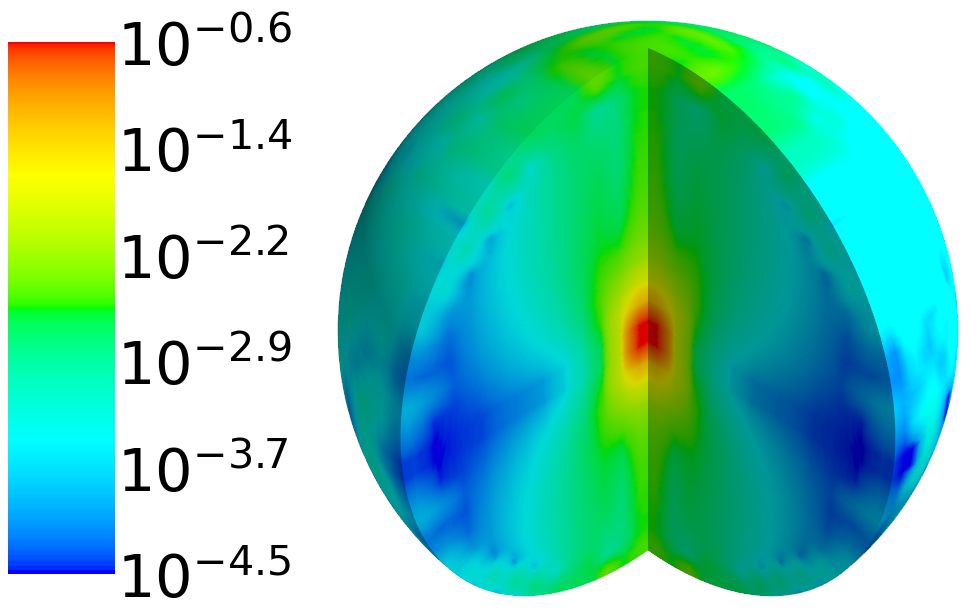}
\caption{The magnitude of Casimir's spin torque, $|\btau_i|$, inside the sphere.}
\label{fig: CST}
\end{figure}

%% file: main.bbl
\begin{thebibliography}{52}%
	\makeatletter
	\providecommand \@ifxundefined [1]{%
		\@ifx{#1\undefined}
	}%
	\providecommand \@ifnum [1]{%
		\ifnum #1\expandafter \@firstoftwo
		\else \expandafter \@secondoftwo
		\fi
	}%
	\providecommand \@ifx [1]{%
		\ifx #1\expandafter \@firstoftwo
		\else \expandafter \@secondoftwo
		\fi
	}%
	\providecommand \natexlab [1]{#1}%
	\providecommand \enquote  [1]{``#1''}%
	\providecommand \bibnamefont  [1]{#1}%
	\providecommand \bibfnamefont [1]{#1}%
	\providecommand \citenamefont [1]{#1}%
	\providecommand \href@noop [0]{\@secondoftwo}%
	\providecommand \href [0]{\begingroup \@sanitize@url \@href}%
	\providecommand \@href[1]{\@@startlink{#1}\@@href}%
	\providecommand \@@href[1]{\endgroup#1\@@endlink}%
	\providecommand \@sanitize@url [0]{\catcode `\\12\catcode `\$12\catcode
		`\&12\catcode `\#12\catcode `\^12\catcode `\_12\catcode `\%12\relax}%
	\providecommand \@@startlink[1]{}%
	\providecommand \@@endlink[0]{}%
	\providecommand \url  [0]{\begingroup\@sanitize@url \@url }%
	\providecommand \@url [1]{\endgroup\@href {#1}{\urlprefix }}%
	\providecommand \urlprefix  [0]{URL }%
	\providecommand \Eprint [0]{\href }%
	\providecommand \doibase [0]{http://dx.doi.org/}%
	\providecommand \selectlanguage [0]{\@gobble}%
	\providecommand \bibinfo  [0]{\@secondoftwo}%
	\providecommand \bibfield  [0]{\@secondoftwo}%
	\providecommand \translation [1]{[#1]}%
	\providecommand \BibitemOpen [0]{}%
	\providecommand \bibitemStop [0]{}%
	\providecommand \bibitemNoStop [0]{.\EOS\space}%
	\providecommand \EOS [0]{\spacefactor3000\relax}%
	\providecommand \BibitemShut  [1]{\csname bibitem#1\endcsname}%
	\let\auto@bib@innerbib\@empty
	\bibitem [{\citenamefont {Weinberg}(2015)}]{Weinberg2015}%
	\BibitemOpen
	\bibfield  {author} {\bibinfo {author} {\bibfnamefont {S.}~\bibnamefont
			{Weinberg}},\ }\href@noop {} {\emph {\bibinfo {title} {Lectures on Quantum
				Mechanics}}}\ (\bibinfo  {publisher} {Cambridge University Press},\ \bibinfo
	{address} {Cambridge},\ \bibinfo {year} {2015})\ pp.\ \bibinfo {pages}
	{xxi--xxii}\BibitemShut {NoStop}%
	\bibitem [{\citenamefont {ge~Chen}\ \emph {et~al.}(2018)\citenamefont
		{ge~Chen}, \citenamefont {Derbes}, \citenamefont {Griffiths}, \citenamefont
		{Hill}, \citenamefont {Sohn},\ and\ \citenamefont {Ting}}]{Coleman2018}%
	\BibitemOpen
	\bibfield  {author} {\bibinfo {author} {\bibfnamefont {B.~G.}\ \bibnamefont
			{ge~Chen}}, \bibinfo {author} {\bibfnamefont {D.}~\bibnamefont {Derbes}},
		\bibinfo {author} {\bibfnamefont {D.}~\bibnamefont {Griffiths}}, \bibinfo
		{author} {\bibfnamefont {B.}~\bibnamefont {Hill}}, \bibinfo {author}
		{\bibfnamefont {R.}~\bibnamefont {Sohn}}, \ and\ \bibinfo {author}
		{\bibfnamefont {Y.-S.}\ \bibnamefont {Ting}},\ }\href {\doibase 10.1142/9371}
	{\emph {\bibinfo {title} {Lectures of Sidney Coleman on Quantum Field
				Theory}}}\ (\bibinfo  {publisher} {{WORLD} {SCIENTIFIC}},\ \bibinfo {year}
	{2018})\BibitemShut {NoStop}%
	\bibitem [{\citenamefont {Mostepanenko}\ and\ \citenamefont
		{Trunov}(1997)}]{Mostepanenko1997}%
	\BibitemOpen
	\bibfield  {author} {\bibinfo {author} {\bibfnamefont {V.~M.}\ \bibnamefont
			{Mostepanenko}}\ and\ \bibinfo {author} {\bibfnamefont {N.~N.}\ \bibnamefont
			{Trunov}},\ }\href@noop {} {\emph {\bibinfo {title} {The Casimir Effect and
				Its Applications}}}\ (\bibinfo  {publisher} {Clarendon Press},\ \bibinfo
	{address} {Oxford, England},\ \bibinfo {year} {1997})\BibitemShut {NoStop}%
	\bibitem [{\citenamefont {Mohideen}\ \emph {et~al.}(2014)\citenamefont
		{Mohideen}, \citenamefont {Klimchitskaya}, \citenamefont {Bordag},\ and\
		\citenamefont {Mostepanenko}}]{Mohideen2014}%
	\BibitemOpen
	\bibfield  {author} {\bibinfo {author} {\bibfnamefont {U.}~\bibnamefont
			{Mohideen}}, \bibinfo {author} {\bibfnamefont {G.~L.}\ \bibnamefont
			{Klimchitskaya}}, \bibinfo {author} {\bibfnamefont {M.}~\bibnamefont
			{Bordag}}, \ and\ \bibinfo {author} {\bibfnamefont {V.}~\bibnamefont
			{Mostepanenko}},\ }\href@noop {} {\emph {\bibinfo {title} {Advances in the
				Casimir Effect}}},\ International Series of Monographs on Physics\ (\bibinfo
	{publisher} {Oxford University Press},\ \bibinfo {address} {London,
		England},\ \bibinfo {year} {2014})\BibitemShut {NoStop}%
	\bibitem [{\citenamefont {Abrikosov}(1975)}]{Abrikosov1975}%
	\BibitemOpen
	\bibfield  {author} {\bibinfo {author} {\bibfnamefont {A.~A.}\ \bibnamefont
			{Abrikosov}},\ }\href@noop {} {\emph {\bibinfo {title} {Methods of quantum
				field theory in statistical physics}}},\ Dover Books on Physics\ (\bibinfo
	{publisher} {Dover Publications},\ \bibinfo {address} {Mineola, NY},\
	\bibinfo {year} {1975})\BibitemShut {NoStop}%
	\bibitem [{\citenamefont {Lamoreaux}(1997)}]{Lamoreaux1997}%
	\BibitemOpen
	\bibfield  {author} {\bibinfo {author} {\bibfnamefont {S.~K.}\ \bibnamefont
			{Lamoreaux}},\ }\href {\doibase 10.1103/PhysRevLett.78.5} {\bibfield
		{journal} {\bibinfo  {journal} {Phys. Rev. Lett.}\ }\textbf {\bibinfo
			{volume} {78}},\ \bibinfo {pages} {5} (\bibinfo {year} {1997})}\BibitemShut
	{NoStop}%
	\bibitem [{\citenamefont {Xu}\ \emph {et~al.}(2022)\citenamefont {Xu},
		\citenamefont {Gao}, \citenamefont {Bang}, \citenamefont {Jacob},\ and\
		\citenamefont {Li}}]{Xu2022}%
	\BibitemOpen
	\bibfield  {author} {\bibinfo {author} {\bibfnamefont {Z.}~\bibnamefont
			{Xu}}, \bibinfo {author} {\bibfnamefont {X.}~\bibnamefont {Gao}}, \bibinfo
		{author} {\bibfnamefont {J.}~\bibnamefont {Bang}}, \bibinfo {author}
		{\bibfnamefont {Z.}~\bibnamefont {Jacob}}, \ and\ \bibinfo {author}
		{\bibfnamefont {T.}~\bibnamefont {Li}},\ }\href {\doibase
		10.1038/s41565-021-01026-8} {\bibfield  {journal} {\bibinfo  {journal}
			{Nature Nanotechnology}\ }\textbf {\bibinfo {volume} {17}},\ \bibinfo {pages}
		{148} (\bibinfo {year} {2022})}\BibitemShut {NoStop}%
	\bibitem [{\citenamefont {Rodriguez}\ \emph {et~al.}(2011)\citenamefont
		{Rodriguez}, \citenamefont {Capasso},\ and\ \citenamefont
		{Johnson}}]{Rodriguez2011}%
	\BibitemOpen
	\bibfield  {author} {\bibinfo {author} {\bibfnamefont {A.~W.}\ \bibnamefont
			{Rodriguez}}, \bibinfo {author} {\bibfnamefont {F.}~\bibnamefont {Capasso}},
		\ and\ \bibinfo {author} {\bibfnamefont {S.~G.}\ \bibnamefont {Johnson}},\
	}\href {\doibase 10.1038/nphoton.2011.39} {\bibfield  {journal} {\bibinfo
			{journal} {Nature Photonics}\ }\textbf {\bibinfo {volume} {5}},\ \bibinfo
		{pages} {211} (\bibinfo {year} {2011})}\BibitemShut {NoStop}%
	\bibitem [{\citenamefont {Yuan}\ \emph {et~al.}(2022)\citenamefont {Yuan},
		\citenamefont {Cao}, \citenamefont {Kamra}, \citenamefont {Duine},\ and\
		\citenamefont {Yan}}]{Yuan20221}%
	\BibitemOpen
	\bibfield  {author} {\bibinfo {author} {\bibfnamefont {H.}~\bibnamefont
			{Yuan}}, \bibinfo {author} {\bibfnamefont {Y.}~\bibnamefont {Cao}}, \bibinfo
		{author} {\bibfnamefont {A.}~\bibnamefont {Kamra}}, \bibinfo {author}
		{\bibfnamefont {R.~A.}\ \bibnamefont {Duine}}, \ and\ \bibinfo {author}
		{\bibfnamefont {P.}~\bibnamefont {Yan}},\ }\href {\doibase
		https://doi.org/10.1016/j.physrep.2022.03.002} {\bibfield  {journal}
		{\bibinfo  {journal} {Physics Reports}\ }\textbf {\bibinfo {volume} {965}},\
		\bibinfo {pages} {1} (\bibinfo {year} {2022})},\ \bibinfo {note} {quantum
		magnonics: When magnon spintronics meets quantum information
		science}\BibitemShut {NoStop}%
	\bibitem [{\citenamefont {Auerbach}(1994)}]{Auerbach1994}%
	\BibitemOpen
	\bibfield  {author} {\bibinfo {author} {\bibfnamefont {A.}~\bibnamefont
			{Auerbach}},\ }\href {\doibase 10.1007/978-1-4612-0869-3} {\emph {\bibinfo
			{title} {Interacting Electrons and Quantum Magnetism}}}\ (\bibinfo
	{publisher} {Springer New York},\ \bibinfo {year} {1994})\BibitemShut
	{NoStop}%
	\bibitem [{\citenamefont {Sachdev}(2011)}]{Sachdev2011}%
	\BibitemOpen
	\bibfield  {author} {\bibinfo {author} {\bibfnamefont {S.}~\bibnamefont
			{Sachdev}},\ }\href {\doibase 10.1017/cbo9780511973765} {\emph {\bibinfo
			{title} {Quantum Phase Transitions}}}\ (\bibinfo  {publisher} {Cambridge
		University Press},\ \bibinfo {year} {2011})\BibitemShut {NoStop}%
	\bibitem [{\citenamefont {Broholm}\ \emph {et~al.}(2020)\citenamefont
		{Broholm}, \citenamefont {Cava}, \citenamefont {Kivelson}, \citenamefont
		{Nocera}, \citenamefont {Norman},\ and\ \citenamefont
		{Senthil}}]{Broholm2020}%
	\BibitemOpen
	\bibfield  {author} {\bibinfo {author} {\bibfnamefont {C.}~\bibnamefont
			{Broholm}}, \bibinfo {author} {\bibfnamefont {R.~J.}\ \bibnamefont {Cava}},
		\bibinfo {author} {\bibfnamefont {S.~A.}\ \bibnamefont {Kivelson}}, \bibinfo
		{author} {\bibfnamefont {D.~G.}\ \bibnamefont {Nocera}}, \bibinfo {author}
		{\bibfnamefont {M.~R.}\ \bibnamefont {Norman}}, \ and\ \bibinfo {author}
		{\bibfnamefont {T.}~\bibnamefont {Senthil}},\ }\href {\doibase
		10.1126/science.aay0668} {\bibfield  {journal} {\bibinfo  {journal}
			{Science}\ }\textbf {\bibinfo {volume} {367}} (\bibinfo {year} {2020}),\
		10.1126/science.aay0668}\BibitemShut {NoStop}%
	\bibitem [{\citenamefont {Rold\'an-Molina}\ \emph {et~al.}(2015)\citenamefont
		{Rold\'an-Molina}, \citenamefont {Santander}, \citenamefont {Nunez},\ and\
		\citenamefont {Fern\'andez-Rossier}}]{Roldan2015}%
	\BibitemOpen
	\bibfield  {author} {\bibinfo {author} {\bibfnamefont {A.}~\bibnamefont
			{Rold\'an-Molina}}, \bibinfo {author} {\bibfnamefont {M.~J.}\ \bibnamefont
			{Santander}}, \bibinfo {author} {\bibfnamefont {A.~S.}\ \bibnamefont
			{Nunez}}, \ and\ \bibinfo {author} {\bibfnamefont {J.}~\bibnamefont
			{Fern\'andez-Rossier}},\ }\href {\doibase 10.1103/PhysRevB.92.245436}
	{\bibfield  {journal} {\bibinfo  {journal} {Phys. Rev. B}\ }\textbf {\bibinfo
			{volume} {92}},\ \bibinfo {pages} {245436} (\bibinfo {year}
		{2015})}\BibitemShut {NoStop}%
	\bibitem [{\citenamefont {Rold\'an-Molina}\ \emph {et~al.}(2014)\citenamefont
		{Rold\'an-Molina}, \citenamefont {Santander}, \citenamefont {N\'u\~nez},\
		and\ \citenamefont {Fern\'andez-Rossier}}]{Roldan2014}%
	\BibitemOpen
	\bibfield  {author} {\bibinfo {author} {\bibfnamefont {A.}~\bibnamefont
			{Rold\'an-Molina}}, \bibinfo {author} {\bibfnamefont {M.~J.}\ \bibnamefont
			{Santander}}, \bibinfo {author} {\bibfnamefont {A.~S.}\ \bibnamefont
			{N\'u\~nez}}, \ and\ \bibinfo {author} {\bibfnamefont {J.}~\bibnamefont
			{Fern\'andez-Rossier}},\ }\href {\doibase 10.1103/PhysRevB.89.054403}
	{\bibfield  {journal} {\bibinfo  {journal} {Phys. Rev. B}\ }\textbf {\bibinfo
			{volume} {89}},\ \bibinfo {pages} {054403} (\bibinfo {year}
		{2014})}\BibitemShut {NoStop}%
	\bibitem [{\citenamefont {Ochoa}\ and\ \citenamefont
		{Tserkovnyak}(2019)}]{Ochoa2019}%
	\BibitemOpen
	\bibfield  {author} {\bibinfo {author} {\bibfnamefont {H.}~\bibnamefont
			{Ochoa}}\ and\ \bibinfo {author} {\bibfnamefont {Y.}~\bibnamefont
			{Tserkovnyak}},\ }\href {\doibase 10.1142/s0217979219300056} {\bibfield
		{journal} {\bibinfo  {journal} {International Journal of Modern Physics B}\
		}\textbf {\bibinfo {volume} {33}},\ \bibinfo {pages} {1930005} (\bibinfo
		{year} {2019})}\BibitemShut {NoStop}%
	\bibitem [{\citenamefont {Psaroudaki}\ and\ \citenamefont
		{Panagopoulos}(2022)}]{Psaroudaki2022}%
	\BibitemOpen
	\bibfield  {author} {\bibinfo {author} {\bibfnamefont {C.}~\bibnamefont
			{Psaroudaki}}\ and\ \bibinfo {author} {\bibfnamefont {C.}~\bibnamefont
			{Panagopoulos}},\ }\href {\doibase 10.1103/PhysRevB.106.104422} {\bibfield
		{journal} {\bibinfo  {journal} {Phys. Rev. B}\ }\textbf {\bibinfo {volume}
			{106}},\ \bibinfo {pages} {104422} (\bibinfo {year} {2022})}\BibitemShut
	{NoStop}%
	\bibitem [{\citenamefont {Psaroudaki}\ \emph {et~al.}(2017)\citenamefont
		{Psaroudaki}, \citenamefont {Hoffman}, \citenamefont {Klinovaja},\ and\
		\citenamefont {Loss}}]{Psaroudaki2017}%
	\BibitemOpen
	\bibfield  {author} {\bibinfo {author} {\bibfnamefont {C.}~\bibnamefont
			{Psaroudaki}}, \bibinfo {author} {\bibfnamefont {S.}~\bibnamefont {Hoffman}},
		\bibinfo {author} {\bibfnamefont {J.}~\bibnamefont {Klinovaja}}, \ and\
		\bibinfo {author} {\bibfnamefont {D.}~\bibnamefont {Loss}},\ }\href {\doibase
		10.1103/PhysRevX.7.041045} {\bibfield  {journal} {\bibinfo  {journal} {Phys.
				Rev. X}\ }\textbf {\bibinfo {volume} {7}},\ \bibinfo {pages} {041045}
		(\bibinfo {year} {2017})}\BibitemShut {NoStop}%
	\bibitem [{\citenamefont {Haller}\ \emph {et~al.}(2022)\citenamefont {Haller},
		\citenamefont {Groenendijk}, \citenamefont {Habibi}, \citenamefont
		{Michels},\ and\ \citenamefont {Schmidt}}]{Haller2022}%
	\BibitemOpen
	\bibfield  {author} {\bibinfo {author} {\bibfnamefont {A.}~\bibnamefont
			{Haller}}, \bibinfo {author} {\bibfnamefont {S.}~\bibnamefont {Groenendijk}},
		\bibinfo {author} {\bibfnamefont {A.}~\bibnamefont {Habibi}}, \bibinfo
		{author} {\bibfnamefont {A.}~\bibnamefont {Michels}}, \ and\ \bibinfo
		{author} {\bibfnamefont {T.~L.}\ \bibnamefont {Schmidt}},\ }\href {\doibase
		10.1103/PhysRevResearch.4.043113} {\bibfield  {journal} {\bibinfo  {journal}
			{Phys. Rev. Res.}\ }\textbf {\bibinfo {volume} {4}},\ \bibinfo {pages}
		{043113} (\bibinfo {year} {2022})}\BibitemShut {NoStop}%
	\bibitem [{\citenamefont {Chen}\ and\ \citenamefont {Ma}(2021)}]{Chen2021}%
	\BibitemOpen
	\bibfield  {author} {\bibinfo {author} {\bibfnamefont {Z.}~\bibnamefont
			{Chen}}\ and\ \bibinfo {author} {\bibfnamefont {F.}~\bibnamefont {Ma}},\
	}\href {\doibase 10.1063/5.0061832} {\bibfield  {journal} {\bibinfo
			{journal} {Journal of Applied Physics}\ }\textbf {\bibinfo {volume} {130}},\
		\bibinfo {pages} {090901} (\bibinfo {year} {2021})},\ \Eprint
	{http://arxiv.org/abs/https://doi.org/10.1063/5.0061832}
	{https://doi.org/10.1063/5.0061832} \BibitemShut {NoStop}%
	\bibitem [{\citenamefont {D{\'{\i}}az}\ \emph {et~al.}(2020)\citenamefont
		{D{\'{\i}}az}, \citenamefont {Hirosawa}, \citenamefont {Klinovaja},\ and\
		\citenamefont {Loss}}]{Diaz2020}%
	\BibitemOpen
	\bibfield  {author} {\bibinfo {author} {\bibfnamefont {S.~A.}\ \bibnamefont
			{D{\'{\i}}az}}, \bibinfo {author} {\bibfnamefont {T.}~\bibnamefont
			{Hirosawa}}, \bibinfo {author} {\bibfnamefont {J.}~\bibnamefont {Klinovaja}},
		\ and\ \bibinfo {author} {\bibfnamefont {D.}~\bibnamefont {Loss}},\ }\href
	{\doibase 10.1103/physrevresearch.2.013231} {\bibfield  {journal} {\bibinfo
			{journal} {Physical Review Research}\ }\textbf {\bibinfo {volume} {2}}
		(\bibinfo {year} {2020}),\ 10.1103/physrevresearch.2.013231}\BibitemShut
	{NoStop}%
	\bibitem [{\citenamefont {Rold{\'{a}}n-Molina}\ \emph
		{et~al.}(2016)\citenamefont {Rold{\'{a}}n-Molina}, \citenamefont {Nunez},\
		and\ \citenamefont {Fern{\'{a}}ndez-Rossier}}]{RoldanMolina2016}%
	\BibitemOpen
	\bibfield  {author} {\bibinfo {author} {\bibfnamefont {A.}~\bibnamefont
			{Rold{\'{a}}n-Molina}}, \bibinfo {author} {\bibfnamefont {A.~S.}\
			\bibnamefont {Nunez}}, \ and\ \bibinfo {author} {\bibfnamefont
			{J.}~\bibnamefont {Fern{\'{a}}ndez-Rossier}},\ }\href {\doibase
		10.1088/1367-2630/18/4/045015} {\bibfield  {journal} {\bibinfo  {journal}
			{New Journal of Physics}\ }\textbf {\bibinfo {volume} {18}},\ \bibinfo
		{pages} {045015} (\bibinfo {year} {2016})}\BibitemShut {NoStop}%
	\bibitem [{\citenamefont {Bouaziz}\ \emph {et~al.}(2020)\citenamefont
		{Bouaziz}, \citenamefont {Iba\~nez Azpiroz}, \citenamefont {Guimar\~aes},\
		and\ \citenamefont {Lounis}}]{Bouaziz2020}%
	\BibitemOpen
	\bibfield  {author} {\bibinfo {author} {\bibfnamefont {J.}~\bibnamefont
			{Bouaziz}}, \bibinfo {author} {\bibfnamefont {J.}~\bibnamefont {Iba\~nez
				Azpiroz}}, \bibinfo {author} {\bibfnamefont {F.~S.~M.}\ \bibnamefont
			{Guimar\~aes}}, \ and\ \bibinfo {author} {\bibfnamefont {S.}~\bibnamefont
			{Lounis}},\ }\href {\doibase 10.1103/PhysRevResearch.2.043357} {\bibfield
		{journal} {\bibinfo  {journal} {Phys. Rev. Res.}\ }\textbf {\bibinfo {volume}
			{2}},\ \bibinfo {pages} {043357} (\bibinfo {year} {2020})}\BibitemShut
	{NoStop}%
	\bibitem [{\citenamefont {Nakata}\ and\ \citenamefont
		{Suzuki}(2023)}]{Nakata2023}%
	\BibitemOpen
	\bibfield  {author} {\bibinfo {author} {\bibfnamefont {K.}~\bibnamefont
			{Nakata}}\ and\ \bibinfo {author} {\bibfnamefont {K.}~\bibnamefont
			{Suzuki}},\ }\href {\doibase 10.1103/PhysRevLett.130.096702} {\bibfield
		{journal} {\bibinfo  {journal} {Phys. Rev. Lett.}\ }\textbf {\bibinfo
			{volume} {130}},\ \bibinfo {pages} {096702} (\bibinfo {year}
		{2023})}\BibitemShut {NoStop}%
	\bibitem [{\citenamefont {Beg}\ \emph {et~al.}(2019)\citenamefont {Beg},
		\citenamefont {Pepper}, \citenamefont {Cort{\'e}s-Ortu{\~n}o}, \citenamefont
		{Atie}, \citenamefont {Bisotti}, \citenamefont {Downing}, \citenamefont
		{Kluyver}, \citenamefont {Hovorka},\ and\ \citenamefont {Fangohr}}]{Beg2019}%
	\BibitemOpen
	\bibfield  {author} {\bibinfo {author} {\bibfnamefont {M.}~\bibnamefont
			{Beg}}, \bibinfo {author} {\bibfnamefont {R.~A.}\ \bibnamefont {Pepper}},
		\bibinfo {author} {\bibfnamefont {D.}~\bibnamefont {Cort{\'e}s-Ortu{\~n}o}},
		\bibinfo {author} {\bibfnamefont {B.}~\bibnamefont {Atie}}, \bibinfo {author}
		{\bibfnamefont {M.-A.}\ \bibnamefont {Bisotti}}, \bibinfo {author}
		{\bibfnamefont {G.}~\bibnamefont {Downing}}, \bibinfo {author} {\bibfnamefont
			{T.}~\bibnamefont {Kluyver}}, \bibinfo {author} {\bibfnamefont
			{O.}~\bibnamefont {Hovorka}}, \ and\ \bibinfo {author} {\bibfnamefont
			{H.}~\bibnamefont {Fangohr}},\ }\href {\doibase 10.1038/s41598-019-44462-2}
	{\bibfield  {journal} {\bibinfo  {journal} {Scientific Reports}\ }\textbf
		{\bibinfo {volume} {9}},\ \bibinfo {pages} {7959} (\bibinfo {year}
		{2019})}\BibitemShut {NoStop}%
	\bibitem [{\citenamefont {Tejo}\ \emph {et~al.}(2021)\citenamefont {Tejo},
		\citenamefont {Heredero}, \citenamefont {Chubykalo-Fesenko},\ and\
		\citenamefont {Guslienko}}]{Tejo2022}%
	\BibitemOpen
	\bibfield  {author} {\bibinfo {author} {\bibfnamefont {F.}~\bibnamefont
			{Tejo}}, \bibinfo {author} {\bibfnamefont {R.~H.}\ \bibnamefont {Heredero}},
		\bibinfo {author} {\bibfnamefont {O.}~\bibnamefont {Chubykalo-Fesenko}}, \
		and\ \bibinfo {author} {\bibfnamefont {K.~Y.}\ \bibnamefont {Guslienko}},\
	}\href {\doibase 10.1038/s41598-021-01175-9} {\bibfield  {journal} {\bibinfo
			{journal} {Scientific Reports}\ }\textbf {\bibinfo {volume} {11}},\ \bibinfo
		{pages} {21714} (\bibinfo {year} {2021})}\BibitemShut {NoStop}%
	\bibitem [{\citenamefont {Zambrano-Rabanal}\ \emph {et~al.}(2023)\citenamefont
		{Zambrano-Rabanal}, \citenamefont {Valderrama}, \citenamefont {Tejo},
		\citenamefont {El{\'{\i}}as}, \citenamefont {Nunez}, \citenamefont
		{Carvalho-Santos},\ and\ \citenamefont {Vidal-Silva}}]{ZambranoRabanal2023}%
	\BibitemOpen
	\bibfield  {author} {\bibinfo {author} {\bibfnamefont {C.}~\bibnamefont
			{Zambrano-Rabanal}}, \bibinfo {author} {\bibfnamefont {B.}~\bibnamefont
			{Valderrama}}, \bibinfo {author} {\bibfnamefont {F.}~\bibnamefont {Tejo}},
		\bibinfo {author} {\bibfnamefont {R.~G.}\ \bibnamefont {El{\'{\i}}as}},
		\bibinfo {author} {\bibfnamefont {A.~S.}\ \bibnamefont {Nunez}}, \bibinfo
		{author} {\bibfnamefont {V.~L.}\ \bibnamefont {Carvalho-Santos}}, \ and\
		\bibinfo {author} {\bibfnamefont {N.}~\bibnamefont {Vidal-Silva}},\ }\href
	{\doibase 10.1038/s41598-023-34167-y} {\bibfield  {journal} {\bibinfo
			{journal} {Scientific Reports}\ }\textbf {\bibinfo {volume} {13}} (\bibinfo
		{year} {2023}),\ 10.1038/s41598-023-34167-y}\BibitemShut {NoStop}%
	\bibitem [{\citenamefont {S{\'a}ez}\ \emph {et~al.}(2022)\citenamefont
		{S{\'a}ez}, \citenamefont {D{\'\i}az}, \citenamefont {Vidal-Silva},
		\citenamefont {Escrig},\ and\ \citenamefont {Vogel}}]{Saez2022}%
	\BibitemOpen
	\bibfield  {author} {\bibinfo {author} {\bibfnamefont {G.}~\bibnamefont
			{S{\'a}ez}}, \bibinfo {author} {\bibfnamefont {P.}~\bibnamefont {D{\'\i}az}},
		\bibinfo {author} {\bibfnamefont {N.}~\bibnamefont {Vidal-Silva}}, \bibinfo
		{author} {\bibfnamefont {J.}~\bibnamefont {Escrig}}, \ and\ \bibinfo {author}
		{\bibfnamefont {E.~E.}\ \bibnamefont {Vogel}},\ }\href {\doibase
		https://doi.org/10.1016/j.rinp.2022.105768} {\bibfield  {journal} {\bibinfo
			{journal} {Results in Physics}\ }\textbf {\bibinfo {volume} {39}},\ \bibinfo
		{pages} {105768} (\bibinfo {year} {2022})}\BibitemShut {NoStop}%
	\bibitem [{\citenamefont {Li}\ \emph {et~al.}(2020)\citenamefont {Li},
		\citenamefont {Pierobon}, \citenamefont {Charilaou}, \citenamefont {Braun},
		\citenamefont {Walet}, \citenamefont {L\"{o}ffler}, \citenamefont {Miles},\
		and\ \citenamefont {Moutafis}}]{Li2020}%
	\BibitemOpen
	\bibfield  {author} {\bibinfo {author} {\bibfnamefont {Y.}~\bibnamefont
			{Li}}, \bibinfo {author} {\bibfnamefont {L.}~\bibnamefont {Pierobon}},
		\bibinfo {author} {\bibfnamefont {M.}~\bibnamefont {Charilaou}}, \bibinfo
		{author} {\bibfnamefont {H.-B.}\ \bibnamefont {Braun}}, \bibinfo {author}
		{\bibfnamefont {N.~R.}\ \bibnamefont {Walet}}, \bibinfo {author}
		{\bibfnamefont {J.~F.}\ \bibnamefont {L\"{o}ffler}}, \bibinfo {author}
		{\bibfnamefont {J.~J.}\ \bibnamefont {Miles}}, \ and\ \bibinfo {author}
		{\bibfnamefont {C.}~\bibnamefont {Moutafis}},\ }\href {\doibase
		10.1103/physrevresearch.2.033006} {\bibfield  {journal} {\bibinfo  {journal}
			{Physical Review Research}\ }\textbf {\bibinfo {volume} {2}} (\bibinfo {year}
		{2020}),\ 10.1103/physrevresearch.2.033006}\BibitemShut {NoStop}%
	\bibitem [{\citenamefont {El{\'{\i}}as}\ and\ \citenamefont
		{Verga}(2011)}]{Elas2011}%
	\BibitemOpen
	\bibfield  {author} {\bibinfo {author} {\bibfnamefont {R.~G.}\ \bibnamefont
			{El{\'{\i}}as}}\ and\ \bibinfo {author} {\bibfnamefont {A.}~\bibnamefont
			{Verga}},\ }\href {\doibase 10.1140/epjb/e2011-20146-6} {\bibfield  {journal}
		{\bibinfo  {journal} {The European Physical Journal B}\ }\textbf {\bibinfo
			{volume} {82}},\ \bibinfo {pages} {159} (\bibinfo {year} {2011})}\BibitemShut
	{NoStop}%
	\bibitem [{\citenamefont {D\"{o}ring}(1968)}]{Dring1968}%
	\BibitemOpen
	\bibfield  {author} {\bibinfo {author} {\bibfnamefont {W.}~\bibnamefont
			{D\"{o}ring}},\ }\href {\doibase 10.1063/1.1656144} {\bibfield  {journal}
		{\bibinfo  {journal} {Journal of Applied Physics}\ }\textbf {\bibinfo
			{volume} {39}},\ \bibinfo {pages} {1006} (\bibinfo {year}
		{1968})}\BibitemShut {NoStop}%
	\bibitem [{\citenamefont {Galkina}\ \emph {et~al.}(1993)\citenamefont
		{Galkina}, \citenamefont {Ivanov},\ and\ \citenamefont
		{Stephanovich}}]{Galkina1993}%
	\BibitemOpen
	\bibfield  {author} {\bibinfo {author} {\bibfnamefont {E.}~\bibnamefont
			{Galkina}}, \bibinfo {author} {\bibfnamefont {B.}~\bibnamefont {Ivanov}}, \
		and\ \bibinfo {author} {\bibfnamefont {V.}~\bibnamefont {Stephanovich}},\
	}\href {\doibase https://doi.org/10.1016/0304-8853(93)90441-4} {\bibfield
		{journal} {\bibinfo  {journal} {Journal of Magnetism and Magnetic Materials}\
		}\textbf {\bibinfo {volume} {118}},\ \bibinfo {pages} {373} (\bibinfo {year}
		{1993})}\BibitemShut {NoStop}%
	\bibitem [{\citenamefont {Izmozherov}\ \emph {et~al.}(2019)\citenamefont
		{Izmozherov}, \citenamefont {Zverev},\ and\ \citenamefont
		{Baykenov}}]{Izmozherov2019}%
	\BibitemOpen
	\bibfield  {author} {\bibinfo {author} {\bibfnamefont {I.~M.}\ \bibnamefont
			{Izmozherov}}, \bibinfo {author} {\bibfnamefont {V.~V.}\ \bibnamefont
			{Zverev}}, \ and\ \bibinfo {author} {\bibfnamefont {E.~Z.}\ \bibnamefont
			{Baykenov}},\ }\href {\doibase 10.1088/1742-6596/1389/1/012002} {\bibfield
		{journal} {\bibinfo  {journal} {Journal of Physics: Conference Series}\
		}\textbf {\bibinfo {volume} {1389}},\ \bibinfo {pages} {012002} (\bibinfo
		{year} {2019})}\BibitemShut {NoStop}%
	\bibitem [{\citenamefont {Hierro-Rodriguez}\ \emph {et~al.}(2020)\citenamefont
		{Hierro-Rodriguez}, \citenamefont {Quir{\'o}s}, \citenamefont {Sorrentino},
		\citenamefont {Alvarez-Prado}, \citenamefont {Mart{\'\i}n}, \citenamefont
		{Alameda}, \citenamefont {McVitie}, \citenamefont {Pereiro}, \citenamefont
		{V{\'e}lez},\ and\ \citenamefont {Ferrer}}]{Hierro-Rodriguez2020}%
	\BibitemOpen
	\bibfield  {author} {\bibinfo {author} {\bibfnamefont {A.}~\bibnamefont
			{Hierro-Rodriguez}}, \bibinfo {author} {\bibfnamefont {C.}~\bibnamefont
			{Quir{\'o}s}}, \bibinfo {author} {\bibfnamefont {A.}~\bibnamefont
			{Sorrentino}}, \bibinfo {author} {\bibfnamefont {L.~M.}\ \bibnamefont
			{Alvarez-Prado}}, \bibinfo {author} {\bibfnamefont {J.~I.}\ \bibnamefont
			{Mart{\'\i}n}}, \bibinfo {author} {\bibfnamefont {J.~M.}\ \bibnamefont
			{Alameda}}, \bibinfo {author} {\bibfnamefont {S.}~\bibnamefont {McVitie}},
		\bibinfo {author} {\bibfnamefont {E.}~\bibnamefont {Pereiro}}, \bibinfo
		{author} {\bibfnamefont {M.}~\bibnamefont {V{\'e}lez}}, \ and\ \bibinfo
		{author} {\bibfnamefont {S.}~\bibnamefont {Ferrer}},\ }\href {\doibase
		10.1038/s41467-020-20119-x} {\bibfield  {journal} {\bibinfo  {journal}
			{Nature Communications}\ }\textbf {\bibinfo {volume} {11}},\ \bibinfo {pages}
		{6382} (\bibinfo {year} {2020})}\BibitemShut {NoStop}%
	\bibitem [{\citenamefont {Andreas}\ \emph {et~al.}(2014)\citenamefont
		{Andreas}, \citenamefont {K\'akay},\ and\ \citenamefont
		{Hertel}}]{Andreas2014}%
	\BibitemOpen
	\bibfield  {author} {\bibinfo {author} {\bibfnamefont {C.}~\bibnamefont
			{Andreas}}, \bibinfo {author} {\bibfnamefont {A.}~\bibnamefont {K\'akay}}, \
		and\ \bibinfo {author} {\bibfnamefont {R.}~\bibnamefont {Hertel}},\ }\href
	{\doibase 10.1103/PhysRevB.89.134403} {\bibfield  {journal} {\bibinfo
			{journal} {Phys. Rev. B}\ }\textbf {\bibinfo {volume} {89}},\ \bibinfo
		{pages} {134403} (\bibinfo {year} {2014})}\BibitemShut {NoStop}%
	\bibitem [{\citenamefont {El\'{\i}as}\ \emph {et~al.}(2014)\citenamefont
		{El\'{\i}as}, \citenamefont {Carvalho-Santos}, \citenamefont {N\'u\~nez},\
		and\ \citenamefont {Verga}}]{Elias2014}%
	\BibitemOpen
	\bibfield  {author} {\bibinfo {author} {\bibfnamefont {R.~G.}\ \bibnamefont
			{El\'{\i}as}}, \bibinfo {author} {\bibfnamefont {V.~L.}\ \bibnamefont
			{Carvalho-Santos}}, \bibinfo {author} {\bibfnamefont {A.~S.}\ \bibnamefont
			{N\'u\~nez}}, \ and\ \bibinfo {author} {\bibfnamefont {A.~D.}\ \bibnamefont
			{Verga}},\ }\href {\doibase 10.1103/PhysRevB.90.224414} {\bibfield  {journal}
		{\bibinfo  {journal} {Phys. Rev. B}\ }\textbf {\bibinfo {volume} {90}},\
		\bibinfo {pages} {224414} (\bibinfo {year} {2014})}\BibitemShut {NoStop}%
	\bibitem [{\citenamefont {Carvalho-Santos}\ \emph {et~al.}(2015)\citenamefont
		{Carvalho-Santos}, \citenamefont {El{\'{\i}}as},\ and\ \citenamefont
		{Nunez}}]{CarvalhoSantos2015}%
	\BibitemOpen
	\bibfield  {author} {\bibinfo {author} {\bibfnamefont {V.}~\bibnamefont
			{Carvalho-Santos}}, \bibinfo {author} {\bibfnamefont {R.}~\bibnamefont
			{El{\'{\i}}as}}, \ and\ \bibinfo {author} {\bibfnamefont {A.~S.}\
			\bibnamefont {Nunez}},\ }\href {\doibase 10.1016/j.aop.2015.10.005}
	{\bibfield  {journal} {\bibinfo  {journal} {Annals of Physics}\ }\textbf
		{\bibinfo {volume} {363}},\ \bibinfo {pages} {364} (\bibinfo {year}
		{2015})}\BibitemShut {NoStop}%
	\bibitem [{\citenamefont {Vansteenkiste}\ \emph {et~al.}(2014)\citenamefont
		{Vansteenkiste}, \citenamefont {Leliaert}, \citenamefont {Dvornik},
		\citenamefont {Helsen}, \citenamefont {Garcia-Sanchez},\ and\ \citenamefont
		{Waeyenberge}}]{Vansteenkiste2014}%
	\BibitemOpen
	\bibfield  {author} {\bibinfo {author} {\bibfnamefont {A.}~\bibnamefont
			{Vansteenkiste}}, \bibinfo {author} {\bibfnamefont {J.}~\bibnamefont
			{Leliaert}}, \bibinfo {author} {\bibfnamefont {M.}~\bibnamefont {Dvornik}},
		\bibinfo {author} {\bibfnamefont {M.}~\bibnamefont {Helsen}}, \bibinfo
		{author} {\bibfnamefont {F.}~\bibnamefont {Garcia-Sanchez}}, \ and\ \bibinfo
		{author} {\bibfnamefont {B.~V.}\ \bibnamefont {Waeyenberge}},\ }\href
	{\doibase 10.1063/1.4899186} {\bibfield  {journal} {\bibinfo  {journal}
			{{AIP} Advances}\ }\textbf {\bibinfo {volume} {4}},\ \bibinfo {pages}
		{107133} (\bibinfo {year} {2014})}\BibitemShut {NoStop}%
	\bibitem [{\citenamefont {d'Albuquerque~e Castro}\ \emph
		{et~al.}(2002)\citenamefont {d'Albuquerque~e Castro}, \citenamefont {Altbir},
		\citenamefont {Retamal},\ and\ \citenamefont {Vargas}}]{Albuquerque2002}%
	\BibitemOpen
	\bibfield  {author} {\bibinfo {author} {\bibfnamefont {J.}~\bibnamefont
			{d'Albuquerque~e Castro}}, \bibinfo {author} {\bibfnamefont {D.}~\bibnamefont
			{Altbir}}, \bibinfo {author} {\bibfnamefont {J.~C.}\ \bibnamefont {Retamal}},
		\ and\ \bibinfo {author} {\bibfnamefont {P.}~\bibnamefont {Vargas}},\ }\href
	{\doibase 10.1103/PhysRevLett.88.237202} {\bibfield  {journal} {\bibinfo
			{journal} {Phys. Rev. Lett.}\ }\textbf {\bibinfo {volume} {88}},\ \bibinfo
		{pages} {237202} (\bibinfo {year} {2002})}\BibitemShut {NoStop}%
	\bibitem [{\citenamefont {Holstein}\ and\ \citenamefont
		{Primakoff}(1940)}]{Holstein1940}%
	\BibitemOpen
	\bibfield  {author} {\bibinfo {author} {\bibfnamefont {T.}~\bibnamefont
			{Holstein}}\ and\ \bibinfo {author} {\bibfnamefont {H.}~\bibnamefont
			{Primakoff}},\ }\href {\doibase 10.1103/PhysRev.58.1098} {\bibfield
		{journal} {\bibinfo  {journal} {Phys. Rev.}\ }\textbf {\bibinfo {volume}
			{58}},\ \bibinfo {pages} {1098} (\bibinfo {year} {1940})}\BibitemShut
	{NoStop}%
	\bibitem [{\citenamefont {Bogoljubov}(1958)}]{Bogoljubov1958}%
	\BibitemOpen
	\bibfield  {author} {\bibinfo {author} {\bibfnamefont {N.~N.}\ \bibnamefont
			{Bogoljubov}},\ }\href {\doibase 10.1007/bf02745585} {\bibfield  {journal}
		{\bibinfo  {journal} {Il Nuovo Cimento}\ }\textbf {\bibinfo {volume} {7}},\
		\bibinfo {pages} {794} (\bibinfo {year} {1958})}\BibitemShut {NoStop}%
	\bibitem [{\citenamefont {Wagner}(1986)}]{Wagner1986}%
	\BibitemOpen
	\bibfield  {author} {\bibinfo {author} {\bibfnamefont {M.}~\bibnamefont
			{Wagner}},\ }\href@noop {} {\emph {\bibinfo {title} {Unitary transformations
				in solid state physics}}},\ Modern Problems in Condensed Matter Science S.\
	(\bibinfo  {publisher} {Elsevier Science},\ \bibinfo {address} {London,
		England},\ \bibinfo {year} {1986})\BibitemShut {NoStop}%
	\bibitem [{\citenamefont {Kittel}(1987)}]{Kittel1987}%
	\BibitemOpen
	\bibfield  {author} {\bibinfo {author} {\bibfnamefont {C.}~\bibnamefont
			{Kittel}},\ }\href@noop {} {\emph {\bibinfo {title} {Quantum theory of
				solids}}},\ \bibinfo {edition} {2nd}\ ed.\ (\bibinfo  {publisher} {John Wiley
		\& Sons},\ \bibinfo {address} {Nashville, TN},\ \bibinfo {year}
	{1987})\BibitemShut {NoStop}%
	\bibitem [{\citenamefont {Colpa}(1978)}]{Colpa1978}%
	\BibitemOpen
	\bibfield  {author} {\bibinfo {author} {\bibfnamefont {J.}~\bibnamefont
			{Colpa}},\ }\href {\doibase https://doi.org/10.1016/0378-4371(78)90160-7}
	{\bibfield  {journal} {\bibinfo  {journal} {Physica A: Statistical Mechanics
				and its Applications}\ }\textbf {\bibinfo {volume} {93}},\ \bibinfo {pages}
		{327} (\bibinfo {year} {1978})}\BibitemShut {NoStop}%
	\bibitem [{\citenamefont {Schwinger}\ \emph {et~al.}(1976)\citenamefont
		{Schwinger}, \citenamefont {Milton}, \citenamefont {Tsai}, \citenamefont
		{DeRaad},\ and\ \citenamefont {Clark}}]{Schwinger1976}%
	\BibitemOpen
	\bibfield  {author} {\bibinfo {author} {\bibfnamefont {J.}~\bibnamefont
			{Schwinger}}, \bibinfo {author} {\bibfnamefont {K.~A.}\ \bibnamefont
			{Milton}}, \bibinfo {author} {\bibfnamefont {W.-Y.}\ \bibnamefont {Tsai}},
		\bibinfo {author} {\bibfnamefont {L.~L.}\ \bibnamefont {DeRaad}}, \ and\
		\bibinfo {author} {\bibfnamefont {D.~C.}\ \bibnamefont {Clark}},\ }\href
	{\doibase 10.1016/0003-4916(76)90020-8} {\bibfield  {journal} {\bibinfo
			{journal} {Annals of Physics}\ }\textbf {\bibinfo {volume} {101}},\ \bibinfo
		{pages} {451} (\bibinfo {year} {1976})}\BibitemShut {NoStop}%
	\bibitem [{\citenamefont {Manton}\ and\ \citenamefont
		{Sutcliffe}(2004)}]{Manton2004}%
	\BibitemOpen
	\bibfield  {author} {\bibinfo {author} {\bibfnamefont {N.}~\bibnamefont
			{Manton}}\ and\ \bibinfo {author} {\bibfnamefont {P.}~\bibnamefont
			{Sutcliffe}},\ }\href {\doibase 10.1017/cbo9780511617034} {\emph {\bibinfo
			{title} {Topological Solitons}}}\ (\bibinfo  {publisher} {Cambridge
		University Press},\ \bibinfo {year} {2004})\BibitemShut {NoStop}%
	\bibitem [{\citenamefont {Weinberg}(2012)}]{Weinberg2012}%
	\BibitemOpen
	\bibfield  {author} {\bibinfo {author} {\bibfnamefont {E.~J.}\ \bibnamefont
			{Weinberg}},\ }\href@noop {} {\emph {\bibinfo {title} {Classical solutions in
				quantum field theory: Solitons and instantons in high energy physics}}},\
	Cambridge monographs on mathematical physics\ (\bibinfo  {publisher}
	{Cambridge University Press},\ \bibinfo {address} {Cambridge, England},\
	\bibinfo {year} {2012})\BibitemShut {NoStop}%
	\bibitem [{\citenamefont {Birch}\ \emph {et~al.}(2021)\citenamefont {Birch},
		\citenamefont {Cort{\'{e}}s-Ortu{\~{n}}o}, \citenamefont {Khanh},
		\citenamefont {Seki}, \citenamefont {{\v{S}}tefan{\v{c}}i{\v{c}}},
		\citenamefont {Balakrishnan}, \citenamefont {Tokura},\ and\ \citenamefont
		{Hatton}}]{Birch2021}%
	\BibitemOpen
	\bibfield  {author} {\bibinfo {author} {\bibfnamefont {M.~T.}\ \bibnamefont
			{Birch}}, \bibinfo {author} {\bibfnamefont {D.}~\bibnamefont
			{Cort{\'{e}}s-Ortu{\~{n}}o}}, \bibinfo {author} {\bibfnamefont {N.~D.}\
			\bibnamefont {Khanh}}, \bibinfo {author} {\bibfnamefont {S.}~\bibnamefont
			{Seki}}, \bibinfo {author} {\bibfnamefont {A.}~\bibnamefont
			{{\v{S}}tefan{\v{c}}i{\v{c}}}}, \bibinfo {author} {\bibfnamefont
			{G.}~\bibnamefont {Balakrishnan}}, \bibinfo {author} {\bibfnamefont
			{Y.}~\bibnamefont {Tokura}}, \ and\ \bibinfo {author} {\bibfnamefont {P.~D.}\
			\bibnamefont {Hatton}},\ }\href {\doibase 10.1038/s42005-021-00675-4}
	{\bibfield  {journal} {\bibinfo  {journal} {Communications Physics}\ }\textbf
		{\bibinfo {volume} {4}} (\bibinfo {year} {2021}),\
		10.1038/s42005-021-00675-4}\BibitemShut {NoStop}%
	\bibitem [{\citenamefont {Li}\ \emph {et~al.}(2018)\citenamefont {Li},
		\citenamefont {Xiao},\ and\ \citenamefont {Chang}}]{Li2018}%
	\BibitemOpen
	\bibfield  {author} {\bibinfo {author} {\bibfnamefont {Y.-M.}\ \bibnamefont
			{Li}}, \bibinfo {author} {\bibfnamefont {J.}~\bibnamefont {Xiao}}, \ and\
		\bibinfo {author} {\bibfnamefont {K.}~\bibnamefont {Chang}},\ }\href
	{\doibase 10.1021/acs.nanolett.8b00492} {\bibfield  {journal} {\bibinfo
			{journal} {Nano Letters}\ }\textbf {\bibinfo {volume} {18}},\ \bibinfo
		{pages} {3032} (\bibinfo {year} {2018})}\BibitemShut {NoStop}%
	\bibitem [{\citenamefont {Zhang}\ \emph {et~al.}(2019)\citenamefont {Zhang},
		\citenamefont {Scully},\ and\ \citenamefont {Agarwal}}]{Zhang2019}%
	\BibitemOpen
	\bibfield  {author} {\bibinfo {author} {\bibfnamefont {Z.}~\bibnamefont
			{Zhang}}, \bibinfo {author} {\bibfnamefont {M.~O.}\ \bibnamefont {Scully}}, \
		and\ \bibinfo {author} {\bibfnamefont {G.~S.}\ \bibnamefont {Agarwal}},\
	}\href {\doibase 10.1103/PhysRevResearch.1.023021} {\bibfield  {journal}
		{\bibinfo  {journal} {Phys. Rev. Res.}\ }\textbf {\bibinfo {volume} {1}},\
		\bibinfo {pages} {023021} (\bibinfo {year} {2019})}\BibitemShut {NoStop}%
	\bibitem [{\citenamefont {Yuan}\ \emph {et~al.}(2020)\citenamefont {Yuan},
		\citenamefont {Zheng}, \citenamefont {Ficek}, \citenamefont {He},\ and\
		\citenamefont {Yung}}]{Yuan2020}%
	\BibitemOpen
	\bibfield  {author} {\bibinfo {author} {\bibfnamefont {H.~Y.}\ \bibnamefont
			{Yuan}}, \bibinfo {author} {\bibfnamefont {S.}~\bibnamefont {Zheng}},
		\bibinfo {author} {\bibfnamefont {Z.}~\bibnamefont {Ficek}}, \bibinfo
		{author} {\bibfnamefont {Q.~Y.}\ \bibnamefont {He}}, \ and\ \bibinfo {author}
		{\bibfnamefont {M.-H.}\ \bibnamefont {Yung}},\ }\href {\doibase
		10.1103/PhysRevB.101.014419} {\bibfield  {journal} {\bibinfo  {journal}
			{Phys. Rev. B}\ }\textbf {\bibinfo {volume} {101}},\ \bibinfo {pages}
		{014419} (\bibinfo {year} {2020})}\BibitemShut {NoStop}%
	\bibitem [{\citenamefont {Azimi~Mousolou}\ \emph {et~al.}(2021)\citenamefont
		{Azimi~Mousolou}, \citenamefont {Liu}, \citenamefont {Bergman}, \citenamefont
		{Delin}, \citenamefont {Eriksson}, \citenamefont {Pereiro}, \citenamefont
		{Thonig},\ and\ \citenamefont {Sj\"oqvist}}]{Azimi2021}%
	\BibitemOpen
	\bibfield  {author} {\bibinfo {author} {\bibfnamefont {V.}~\bibnamefont
			{Azimi~Mousolou}}, \bibinfo {author} {\bibfnamefont {Y.}~\bibnamefont {Liu}},
		\bibinfo {author} {\bibfnamefont {A.}~\bibnamefont {Bergman}}, \bibinfo
		{author} {\bibfnamefont {A.}~\bibnamefont {Delin}}, \bibinfo {author}
		{\bibfnamefont {O.}~\bibnamefont {Eriksson}}, \bibinfo {author}
		{\bibfnamefont {M.}~\bibnamefont {Pereiro}}, \bibinfo {author} {\bibfnamefont
			{D.}~\bibnamefont {Thonig}}, \ and\ \bibinfo {author} {\bibfnamefont
			{E.}~\bibnamefont {Sj\"oqvist}},\ }\href {\doibase
		10.1103/PhysRevB.104.224302} {\bibfield  {journal} {\bibinfo  {journal}
			{Phys. Rev. B}\ }\textbf {\bibinfo {volume} {104}},\ \bibinfo {pages}
		{224302} (\bibinfo {year} {2021})}\BibitemShut {NoStop}%
	\bibitem [{\citenamefont {Ren}\ \emph {et~al.}(2022)\citenamefont {Ren},
		\citenamefont {Xie}, \citenamefont {Li}, \citenamefont {Ma},\ and\
		\citenamefont {Li}}]{Ren2022}%
	\BibitemOpen
	\bibfield  {author} {\bibinfo {author} {\bibfnamefont {Y.-l.}\ \bibnamefont
			{Ren}}, \bibinfo {author} {\bibfnamefont {J.-k.}\ \bibnamefont {Xie}},
		\bibinfo {author} {\bibfnamefont {X.-k.}\ \bibnamefont {Li}}, \bibinfo
		{author} {\bibfnamefont {S.-l.}\ \bibnamefont {Ma}}, \ and\ \bibinfo {author}
		{\bibfnamefont {F.-l.}\ \bibnamefont {Li}},\ }\href {\doibase
		10.1103/PhysRevB.105.094422} {\bibfield  {journal} {\bibinfo  {journal}
			{Phys. Rev. B}\ }\textbf {\bibinfo {volume} {105}},\ \bibinfo {pages}
		{094422} (\bibinfo {year} {2022})}\BibitemShut {NoStop}%
\end{thebibliography}
